\documentclass[
superscriptaddress,aip,apl,reprint,twocolumn,amsmath,amssymb,showpacs]{revtex4-2}
\usepackage{}

\usepackage[dvipsnames]{xcolor} 

\usepackage{graphicx}  
\usepackage{mathptmx} 
\usepackage{epstopdf}
\usepackage{dcolumn}   
\usepackage{bm}        
\usepackage{amssymb}   
\usepackage{amsmath}   
\usepackage{amsthm}
\usepackage{chemarrow}
\usepackage{color}
\usepackage{mathrsfs}
\usepackage{float}
\usepackage{bbold}
\usepackage{bbm}
\usepackage[a4paper,colorlinks=true,
linkcolor=blue,citecolor=blue,
pdfauthor={ },
pdftitle={ },
pdfsubject={ },
pdfkeywords={ }]{hyperref}

\theoremstyle{plain}
\newtheorem*{theorem*}{Theorem}

\begin{document}


\title{Search for axion-like dark matter with spin-based amplifiers}

\date{\today}

\author{Min Jiang}
\email[]{These authors contributed equally to this work}
\affiliation{
Hefei National Laboratory for Physical Sciences at the Microscale and Department of Modern Physics, University of Science and Technology of China, Hefei 230026, China}
\affiliation{
CAS Key Laboratory of Microscale Magnetic Resonance, University of Science and Technology of China, Hefei 230026, China}
\affiliation{
Synergetic Innovation Center of Quantum Information and Quantum Physics, University of Science and Technology of China, Hefei 230026, China}

\author{Haowen Su}
\email[]{These authors contributed equally to this work}
\affiliation{
Hefei National Laboratory for Physical Sciences at the Microscale and Department of Modern Physics, University of Science and Technology of China, Hefei 230026, China}
\affiliation{
CAS Key Laboratory of Microscale Magnetic Resonance, University of Science and Technology of China, Hefei 230026, China}
\affiliation{
Synergetic Innovation Center of Quantum Information and Quantum Physics, University of Science and Technology of China, Hefei 230026, China}

\author{Antoine Garcon}
\affiliation{Helmholtz-Institut, GSI Helmholtzzentrum f{\"u}r Schwerionenforschung, Mainz 55128, Germany}
\affiliation{Johannes Gutenberg University, Mainz 55128, Germany}

\author{Xinhua Peng}
\email[]{xhpeng@ustc.edu.cn}
\affiliation{
Hefei National Laboratory for Physical Sciences at the Microscale and Department of Modern Physics, University of Science and Technology of China, Hefei 230026, China}
\affiliation{
CAS Key Laboratory of Microscale Magnetic Resonance, University of Science and Technology of China, Hefei 230026, China}
\affiliation{
Synergetic Innovation Center of Quantum Information and Quantum Physics, University of Science and Technology of China, Hefei 230026, China}

\author{Dmitry Budker}
\affiliation{Helmholtz-Institut, GSI Helmholtzzentrum f{\"u}r Schwerionenforschung, Mainz 55128, Germany}
\affiliation{Johannes Gutenberg University, Mainz 55128, Germany}
\affiliation{Department of Physics, University of California, Berkeley, CA 94720-7300, USA}

\begin{abstract}{
Ultralight axion-like particles (ALPs) are well-motivated dark matter candidates introduced by theories beyond the standard model.
However, the constraints on the existence of ALPs through existing laboratory experiments are hindered by their current sensitivities, which are usually weaker than astrophysical limits.
Here, we demonstrate a new quantum sensor to search for ALPs in the mass range that spans about two decades from 8.3~feV to 744~feV.
Our sensor makes use of hyperpolarized long-lived nuclear spins as a pre-amplifier that effectively enhances coherently oscillating axion-like dark-matter field by a factor of $>$100.
Using spin-based amplifiers,
we achieve an ultrahigh magnetic sensitivity of 18~fT/Hz$^{1/2}$, which is significantly better than state-of-the-art nuclear-spin magnetometers.
Our experiment constrains the parameter space describing the coupling of ALPs to nucleons over our mass range,
at 67.5~feV reaching $2.9\times 10^{-9}~\textrm{GeV}^{-1}$ ($95\%$ confidence level),
improving over previous laboratory limits by at least five orders of magnitude.
Our measurements also constrain the ALP-nucleon quadratic interaction and dark photon-nucleon interaction with new limits beyond the astrophysical ones.}
\end{abstract}

\maketitle





Despite astrophysical evidence for the existence of dark matter, direct detection of its interaction with standard model particles has not been achieved~\cite{bertone2018history}.
Illuminating dark matter is the best hope of making progress in understanding our universe and would provide insights into astrophysics, cosmology and physics beyond the standard model~\cite{demille2017probing, safronova2018search}.
There is a broad range of particle candidates for dark matter~\cite{bertone2018new}.
Weakly interacting massive particles (WIMPs) have attracted the most attention over the past four decades.
Despite many experiments of increasing sensitivity,
there are no undisputed signatures of WIMP existence,
and fundamental background from the neutrino floor will soon limit the detection sensitivity of WIMP searches~\cite{ambrosi2017direct, aprile2017first, liu2017current}.
The axion, emerging from a solution to the strong-CP problem~\cite{peccei1977cp, peccei1977constraints}, is another well-motivated dark matter candidate~\cite{preskill1983cosmology, kim2010axions}.
Since the original concept of axion was proposed,
axions and other light pseudoscalar bosons (collectively referred to as axion-like particles or ALPs)
naturally emerge when a global symmetry is broken at higher energy scale,
for example, in grand unified theory, string theory, and models with extra dimensions~\cite{irastorza2018new, svrcek2006axions}.

Traditional particle-physics techniques such as observation of particle collisions are completely inadequate for ALP searches with light quanta~\cite{demille2017probing,safronova2018search}.
Experimental searches for axion-like dark matter are based on their nongravitational interactions with standard model particles and fields
\cite{demille2017probing,safronova2018search, anastassopoulos2017new, bradley2003microwave, zhong2018results, braine2020extended, ouellet2019first, gramolin2020search, budker2014proposal, roberts2014limiting, graham2011axion, stadnik2014axion, kimball2020overview,jiang2019floquet}.
Recently, a variety of works have focused their searches on axion-nucleon interactions~\cite{abel2017search, kimball2020overview, wu2019search,jiang2019floquet, smorra2019direct, garcon2019constraints, budker2014proposal, graham2018spin, graham2013new, aybas2021search, bloch2020axion};
in this case, ALPs act as a time-oscillating magnetic field which couples to nuclear spins.
Experiments with nuclear magnetic resonance (NMR) techniques directly search for axion-nucleon interactions~\cite{budker2014proposal,kimball2020overview,graham2013new,aybas2021search}.
Several searches for ALPs in the mass range from $10^{-22}$~to~10$^{-17}$eV have measured the energy shift of nuclear spins
and placed stringent limits on the coupling of ALPs and nucleons~\cite{abel2017search, wu2019search, smorra2019direct}.
The Cosmic Axion Spin Precession Experiment (CASPEr) has recently applied zero- to ultralow-field NMR to explore ALPs with masses ranging from $10^{-16}$~eV to $7.8\times 10^{-14}$~eV~[\onlinecite{garcon2019constraints}].
The precision measurement of $^{207}$Pb solid-state NMR has placed a new limit for ALP-nucleon interactions in the mass range $1.62 \times 10^{-7}$~eV to $1.66 \times 10^{-7}$~eV~[\onlinecite{aybas2021search}].
Although limits have been placed on the axion-nucleon coupling, these are still weaker than astrophysical survey limits and a broad mass range is left to be probed.

\begin{figure*}[t]  
	\makeatletter
\centering
	\def\@captype{figure}
	\makeatother
	\includegraphics[scale=0.62]{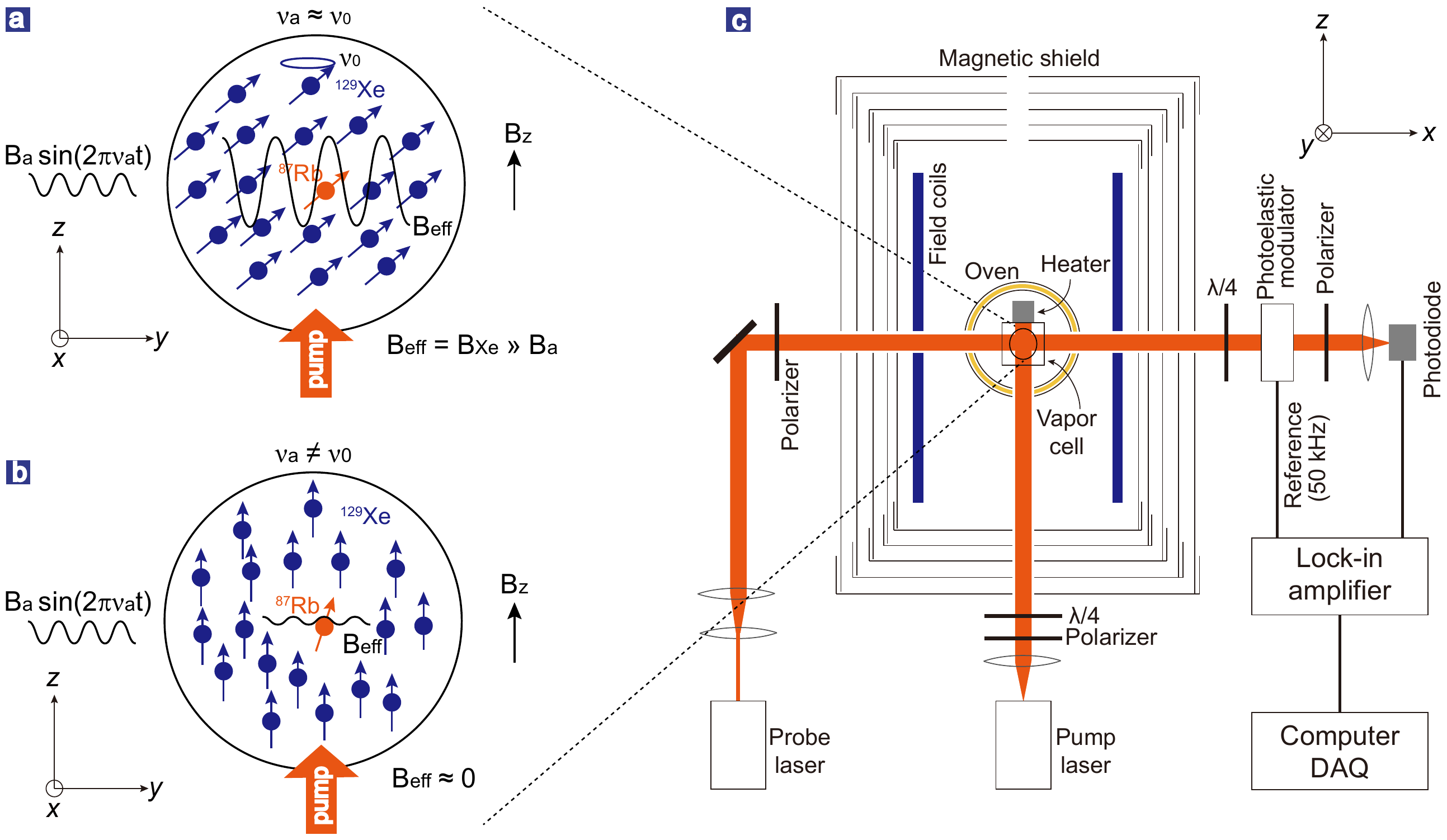}
	\caption{\textbf{Basic principle of the spin-based amplifier}. \textbf{a} and \textbf{b}, Schematic of the spin amplifier based on hyperpolarized long-lived $^{129}$Xe nuclear spins. The key element of the amplifier is a 0.5~cm$^3$ cubic vapor cell containing 5~torr isotopically enriched $^{129}$Xe, 250~torr N$_2$, and a droplet (several milligrams) of isotopically enriched $^{87}$Rb. $^{129}$Xe spins are polarized through spin-exchange collisions with optically polarized $^{87}$Rb atoms. A bias magnetic field  is applied along $z$ to tune the $^{129}$Xe Larmor frequency. When axion-like dark matter field (or generally, external oscillating magnetic field) is resonant with $^{129}$Xe spins (\textbf{a}), the $^{129}$Xe spin magnetization is tilted away from the bias-field direction and their transverse component generates an effective field $\textbf{B}_{\textrm{eff}}$ on $^{87}$Rb atoms. Due to Fermi-contact enhancement, the amplification factor defined as $\eta=|\textbf{B}_{\textrm{eff}} /\textbf{B}_{a}|$ can be significantly larger than one, enabling to greatly amplify the signal from axion-like dark matter field. When the oscillating field is far-off-resonant with $^{129}$Xe (\textbf{b}), $^{129}$Xe spin magnetization remains unchanged and thus $\textbf{B}_{\textrm{eff}}\approx 0$. \textbf{c}, Schematic of $^{87}$Rb magnetometer as a detector of the $^{129}$Xe effective oscillating field $\textbf{B}_{\textrm{eff}}$. The $^{129}$Xe-$^{87}$Rb vapor cell is heated to $140$~$^\circ$C and shielded with a five-layer cylindrical $\mu$-metal shield. In the $^{129}$Xe-$^{87}$Rb vapor cell, $^{87}$Rb atoms are pumped with a circularly polarized laser light tuned to the D1 transition at $795$~nm. The transverse angular momentum of $^{87}$Rb atoms along the ${x}$ axis is measured using a linearly polarized probe light blue-detuned $110$~GHz from the D2 transition at $780$~nm. The $^{87}$Rb transverse angular momentum is sensitive to magnetic fields along  $x$ and $y$ and thus function as a magnetometer (see Supplementary Information), by which the effective magnetic field $\textbf{B}_{\textrm{eff}}$ generated by $^{129}$Xe transverse magnetization can be measured.}
	\label{figure1}
\end{figure*}

\begin{figure*}[t]  
	\makeatletter
\centering
	\def\@captype{figure}
	\makeatother
	\includegraphics[scale=1.8]{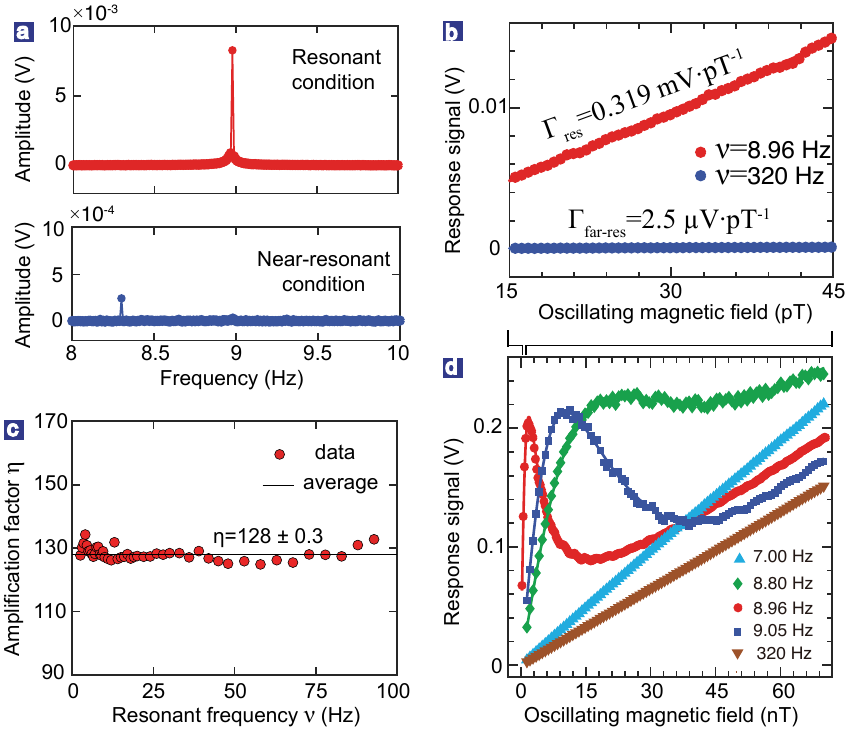}
	\caption{\textbf{Proof-of-principle demonstrations of the spin-based amplifier}. \textbf{a}, The amplification performance of the $^{129}$Xe spin-based amplifier on resonance (top) and near-resonance (bottom). The fast Fourier transformation spectra in the case of $\nu=8.96$ (resonance case), $8.3$~Hz (near-resonance case) are shown in the top and bottom of \textbf{a}, respectively. \textbf{b}, The linear response signals (red circles, on-resonance; blue circles, far off-resonance) as a function of the oscillating field amplitude from 15~pT to 45~pT. The on-resonance slope $\Gamma_{\textrm{on-res}}\approx 0.319$~mV$\cdot$pT$^{-1}$ is at least two orders of magnitude greater than the far-off-resonance slope $\Gamma_{\textrm{far-res}} \approx 2.5$~$\mu$V$\cdot$pT$^{-1}$. \textbf{c}, The amplification factor $\eta$ as a function of resonant frequency (corresponding to different applied bias fields). Experimental data shown in red circles yield their mean $\eta = 128 \pm 0.3$. \textbf{d}, The nonlinear responses as a function of the oscillating field amplitude over a large region (below 70~nT). In the near-zero-amplitude regime, the response of the spin-based amplifier is linear, corresponding to the case shown in \textbf{b}. In the experiments of \textbf{a}, \textbf{b} and \textbf{d}, the bias field is set at $B_z\approx 759$~nT, corresponding to $^{129}$Xe Larmor frequency $\nu_0 \approx 8.96$~Hz.}
	\label{figure2}
\end{figure*}

\begin{figure}[t]  
	\makeatletter
\centering
	\def\@captype{figure}
	\makeatother
	\includegraphics[scale=1.35]{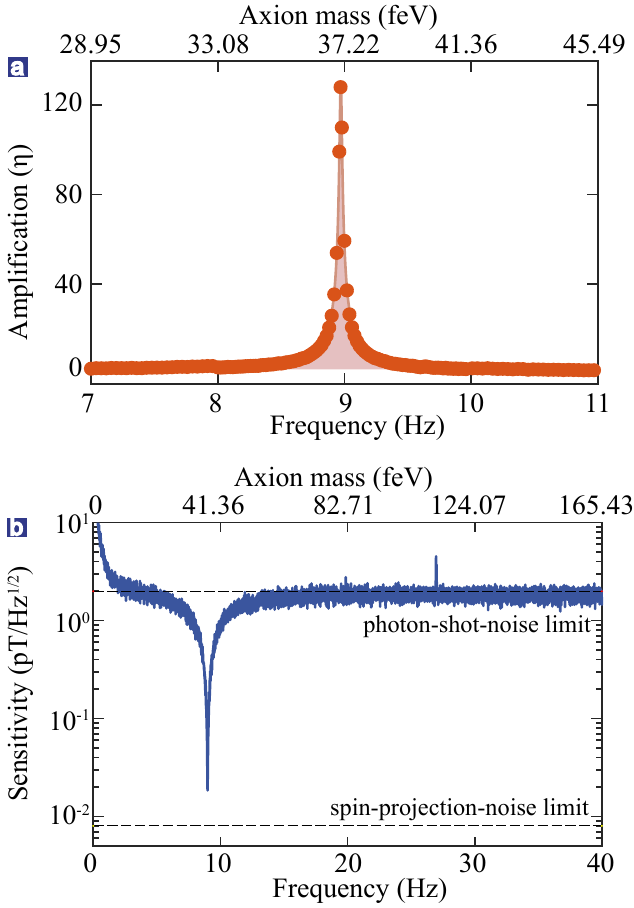}
	\caption{\textbf{Amplification and sensitivity of the magnetic field measurement assisted with the spin-based amplifier}. \textbf{a}, The frequency-response of the $^{87}$Rb magnetometer to oscillating 30-pT $y$-fields, assisted with $^{129}$Xe spin-based amplifier. The experimental data (red circles) are obtained by scanning the oscillating field frequencies. The solid line is the theoretical fit of the data and agrees well with the experiment. The fit yields a bandwidth $\approx 0.052$~Hz of the spin-based amplifier. \textbf{b}, The magnetic sensitivity of the spin-amplifier-based magnetometer. Note that $18~\textrm{fT} / \textrm{Hz}^{1/2}$ is achieved at $^{129}$Xe Larmor frequency, which is beyond the photon-shot-noise limit and comparable to the spin-projection-noise limit of the rubidium magnetometer itself.}
	\label{figure3}
\end{figure}

\begin{figure*}[t]  
	\makeatletter
\centering
	\def\@captype{figure}
	\makeatother
	\includegraphics[scale=1.6]{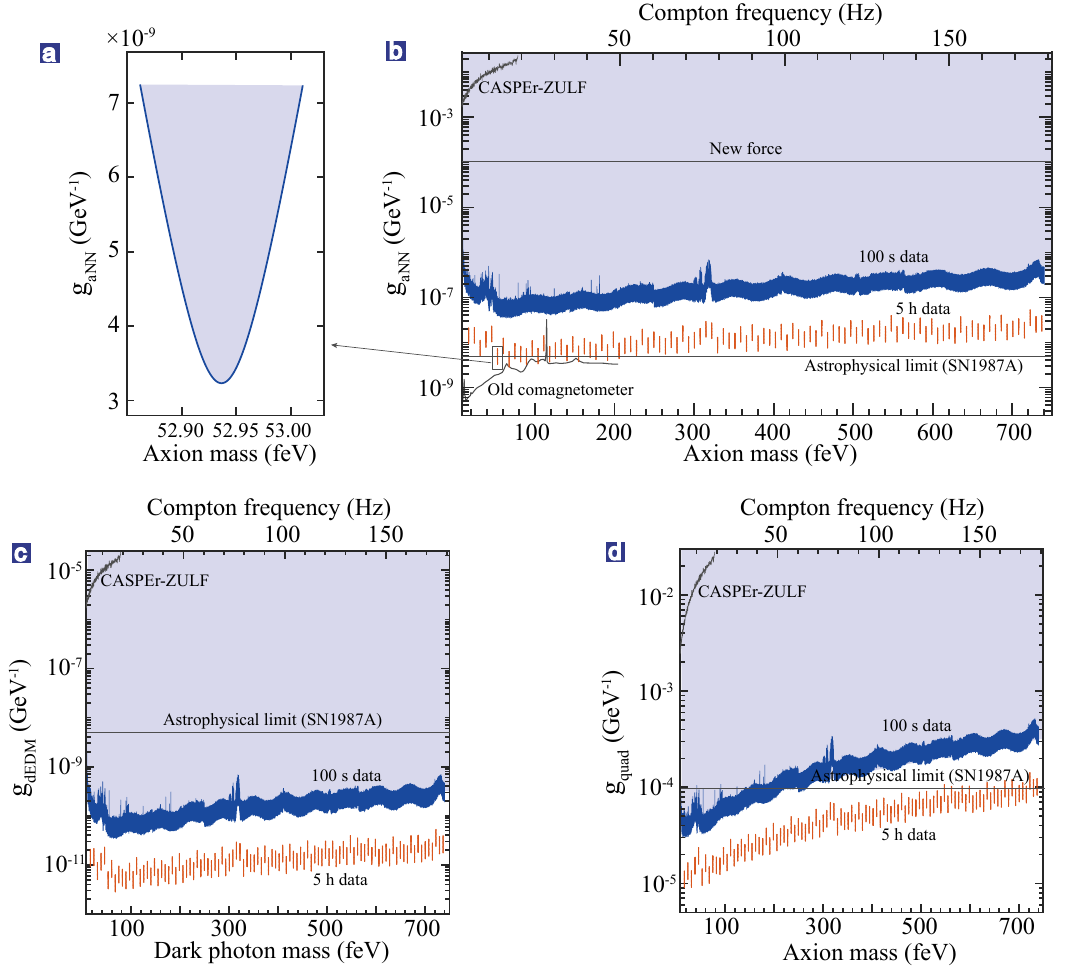}
	\caption{\textbf{Results of axion-like dark matter search}. \textbf{a}, Limits on the coupling strength of nucleons with the axion-like dark matter within a mass range centred at $m_a\approx 52.94$~feV and with the width of 0.15~feV. The blue-shaded region ($95\%$ confidence level) is excluded by our five-hour measurement at $\nu_0 \approx 12.80$~Hz of the spin-based amplifier. \textbf{b}, Limits on the coupling strength of nucleons with the axion-like dark matter in the mass range from 8.3~feV to 744~feV. The blue-shade is excluded by our measurements (100~s for each run) at the $95\%$ confidence level. The red lines show our advanced sensitivity (five hours for each run) for axion-like dark matter at 100 probed axion masses with a window width of 0.15~feV, as shown in \textbf{a}. The grey line shows the limit given by the CASPEr-ZULF experiment~\cite{garcon2019constraints}. The ``Old comagnetometer'' limit comes from ref.~[\onlinecite{bloch2020axion}], which analyses the old comagnetometer data from previously published work. The horizontal black lines show the laboratory searches for new spin-dependent forces and the astrophysical limit from supernova SN1987A cooling~\cite{vysotsskii1978some, raffelt2008astrophysical}. Limits on dark photon-nucleon coupling (\textbf{c}) and axion-like quadratic coupling with nucleon (\textbf{d}) in the mass range from 8.3~feV to 744~feV. The blue-shade is excluded by our measurements (100~s for each run) at the $95\%$ confidence level. The red lines show our advanced sensitivity (five hours for each run) at 100 probed masses with a window width of 0.15~feV. The grey line shows the limit given by the CASPEr-ZULF experiment~\cite{garcon2019constraints}. The horizontal black line shows the astrophysical limit from supernova SN1987A cooling~\cite{vysotsskii1978some, raffelt2008astrophysical}.}
	\label{figure4}
\end{figure*}



In this article, we describe a realization of a new quantum sensor assisted with a spin-based amplifier to search for ALP dark matter,
with which the search sensitivity is comparable to and even beyond the stringent astrophysical survey limits~\cite{vysotsskii1978some, raffelt2008astrophysical}.
The key new ingredient is the use of hyperpolarized long-lived nuclear spins as a pre-amplifier to significantly amplify the signal from coherently oscillating ALP dark matter field by a factor of $>$100.
Then the amplified ALP signal can be searched with a conventional atomic magnetometer~\cite{kominis2003subfemtotesla, budker2007optical}.
We would like to emphasize the difference of ALP search approaches between this work and other searches.
Resonant NMR searches for ALPs have been proposed~\cite{budker2014proposal, graham2018spin, graham2013new, arvanitaki2014resonantly} and their experimental demonstrations are ongoing.
Such works all consider the situation where the nuclear spins are measured from their proximity with atomic and SQUIDs magnetometers;
in this case, it is experimentally challenging to prepare nuclear spins with high spin polarization and maintain readout sensitivity.
Unlike previous works, our work uses a different scheme in which nuclear spins and the detector spatially overlap in a same vapor cell,
offering two significant advantages:
nuclear spins can be directly hyperpolarized to achieve a polarization of 0.1-0.3 by spin-exchange optical pumping~\cite{walker1997spin,jiang2019floquet}, significantly better than recently demonstrated thermal polarization $\sim 10^{-6}$~[\onlinecite{wu2019search}, \onlinecite{garcon2019constraints}];
nuclear spin signals can be enhanced due to large Fermi-contact enhancement factor~\cite{walker1997spin, jiang2019floquet}, detected $in$ $situ$ with an atomic magnetometer.
For the heavy noble gas, such as $^{129}$Xe, the magnetic field generated by nuclear magnetization can be enhanced by a large factor of 540.
In this way we realize the spin-based amplification of ALP signals and immediately achieve high search sensitivity to ALPs.



Axion-like particles could have been produced in the early Universe by nonthermal mechanisms,
such as ``vacuum misalignment''~\cite{marsh2016axion}.
They form a classical field $a\approx a_0 \cos (2 \pi \nu_a t)$~[\onlinecite{kimball2020overview, abel2017search, wu2019search, smorra2019direct, garcon2019constraints, graham2018spin, graham2013new}],
oscillating at its Compton frequency $\nu_a=m_a c^2/h$,
where $m_a$ is the axion mass, $c$ is the  speed of light, and $h$ is the Planck constant.
In the standard halo model, the characteristic coherence time of this oscillating $\textrm{ALPs}$ field is $\sim 10^6$ periods.
The field amplitude $a_0$ can be estimated by the galactic dark-matter energy density $\rho_a=m_a^2 a^2_0 c^2/(2 \hbar^2)\approx 0.4$~$\textrm{GeV}~\textrm{cm}^{-3}$
[\onlinecite{preskill1983cosmology}, \onlinecite{dine1983not}].
ALPs would interact with nuclear spins by a Zeeman-like Hamiltonian $H_{\textrm{int}}\approx \gamma  \textbf{B}_a \cdot \mathbf{I}_N$,
where $\textbf{B}_a= \textrm{g}_{\textrm{aNN}} \sqrt{2 \hbar c \rho_a} \sin(2\pi\nu_a t) \bm{v}_a/\gamma$ represents the effective magnetic field induced by ALP dark matter~\cite{kimball2020overview, abel2017search, wu2019search, smorra2019direct, garcon2019constraints} (see Supplementary Information)
and $\mathbf{I}_N$ is the nuclear spin.
Here, $\textrm{g}_{\textrm{aNN}}$ is the strength of the axion-nucleon coupling,
$|\bm{v}_a| \sim 10^{-3}c$ represents the local galactic virial velocity~\cite{kimball2020overview},
and $\gamma$ is the gyromagnetic ratio of the nuclear spin.
Our ALP search scheme is based on detecting the effective $\textbf{B}_a$ and measuring or constraining the strength $\textrm{g}_{\textrm{aNN}}$ over range of ALP masses.

Our approach involves the use of hyperpolarized $^{129}$Xe gas as a spin-based amplifier.
$^{129}$Xe spatially overlaps with $^{87}$Rb in the same vapor cell,
where $^{129}$Xe nuclear spins act as an ALP pre-amplifier and $^{87}$Rb magnetometer further reads out the amplified ALP signal.
When the ALP Compton frequency ($\nu_a$) matches the $^{129}$Xe spin Larmor frequency ($\nu_0$$=$$\gamma B_z$), i.e., $\nu_a \approx \nu_0$,
$\textrm{ALPs}$ tilt $^{129}$Xe spins away from the direction of the applied bias magnetic field,
as shown in Fig.~\ref{figure1}\textbf{a}.
Subsequently, the detection signature of ALP dark matter is an oscillating nuclear transverse magnetization $\textbf{M}_{n}(t)$ (see Supplementary Information).
The pre-amplified effective field experienced by the $^{87}$Rb atoms is given by $\textbf{B}_{\textrm{eff}}= (8 \pi \kappa_{0}/3) \textbf{M}_n$~[\onlinecite{jiang2019floquet}, \onlinecite{walker1997spin}],
which is read out by $^{87}$Rb magnetometer.
Due to the large Fermi-contact enhancement factor $\kappa_{0} \approx 540$ between the $^{87}$Rb and $^{129}$Xe,
the effective field $\textbf{B}_{\textrm{eff}}$ can be significantly larger than the ALP field ($|\textbf{B}_{\textrm{eff}}| \gg |\textbf{B}_a|$).
This allows us to achieve considerable amplitude amplification of the ALP signal and therefore improve sensitivity to ALP dark matter.
In contrast, when the ALP frequency is different from the Larmor frequency,
there is no measurable transverse magnetization $\textbf{M}_n(t)\approx 0$ and thus $\textbf{B}_{\textrm{eff}}\approx 0$ (Fig.~\ref{figure1}\textbf{b}).

We experimentally measure the amplification of the signal from ALP field through applying a weak oscillating magnetic field to simulate an ALP field.
The amplification factor is defined as $\eta=|\textbf{B}_{\textrm{eff}}/\textbf{B}_a|$.
For example, the bias field is set at $B_z \approx 759$~nT, corresponding to $\nu_0 \approx 8.96$~Hz.
A simulated ALP oscillating field with amplitude $\approx30$~pT and frequency $8.96$~Hz or $8.30$~Hz is applied to the vapor cell along ${y}$.
We collect $100$~s of data and then perform a discrete Fourier transform on the acquired data.
As shown in the top of Fig.~\ref{figure2}\textbf{a},
the signal is greatly enhanced when the applied oscillating field frequency coincides with the $^{129}$Xe Larmor frequency.
This is in contrast to the near-resonant condition (Fig.~\ref{figure2}\textbf{a}, bottom),
where the signal is much weaker.
Figure~\ref{figure2}\textbf{b} shows that both on-resonant and far-off-resonant ($320$~Hz) signals linearly depend on the magnitude of oscillating fields ranging from $15$ to $45$~pT.
The slope of the response signal at resonance is $\Gamma_{\textrm{res}} \approx 0.319$~mV/pT,
which represents an improvement by a factor of $\eta \approx 127.6$ over $\Gamma_{\textrm{far-res}} \approx 2.5$~$\mu$V/pT of far-off-resonance case.
This indicates that the external oscillating field is effectively pre-amplified from 30 pT to $|\textbf{B}_{\textrm{eff}} |\approx 4080$~pT.
Figure~\ref{figure2}\textbf{c} further shows amplification factors $\eta$ as a function of resonant frequencies tuned by bias fields,
which yield the mean $\eta= 128 \pm 0.3$ (black line) (Supplementary Information).

With increasing the amplitude of oscillating fields,
we find that the response of the spin-based amplifier to oscillating fields becomes nonlinear.
Figure~\ref{figure2}\textbf{d} provides experimental nonlinear response signals, scanned as a function of the oscillating field amplitude.
In contrast to the far-off-resonant cases (7 Hz, 320 Hz),
the signals for resonant ($8.96$~Hz) and near-resonant cases ($8.80$~Hz, $9.05$~Hz) first increase, then decrease, and lastly increase again,
originating from nuclear magnetization saturation.
The solid curves in Fig.~\ref{figure2}\textbf{d} represent our theoretical calculations (Supplementary Information), which agree well with the experiment.
A particularly sensitive window for measuring external oscillating fields corresponds to the amplitude of oscillating fields below several nT ($B_{\textrm{a}} \ll \frac{2}{\gamma \sqrt{T_1 T_2}}$, where $T_{1,2}$ is longitudinal and transverse spin relaxation time, respectively; see Supplementary Information).
In practice, we are interested in a sensitive measurement of small magnetic fields (for example, axion-like dark matter field),
whose amplitudes are naturally within the sensitive window demonstrated in this work.
To maintain our ALP search in the sensitive window,
we carefully suppress ambient electromagnetic interference, for example, by using a five-layer magnetic shield and an ultralow-noise current source.

Having established the sensing technique,
we now quantify the detection sensitivity of our experiment to an ALP dark matter signal.
We first calibrate the frequency dependence of our sensor by scanning the frequencies of the oscillating fields and recording the corresponding amplification factor.
For example, in a bias field of $759$~nT,
the sensor amplification $\eta$ is illustrated in a small frequency range around the resonant frequency ($\nu_0 \approx 8.96$~Hz).
It reaches maximum on resonance (Fig.~\ref{figure3}\textbf{a});
the frequency dependence of the response is consistent with a single-pole band-pass filter model with a full-width at half-maximum (FWHM) of $0.052$~Hz (Supplementary Information).
The FWHM value corresponds to the axion mass range of 0.22~feV.
Using such a spin-based amplifier that enhances the measured oscillating field,
we achieve a magnetic sensitivity of $18~\textrm{fT} / \textrm{Hz}^{1/2}$ at the resonant frequency in Fig.~\ref{figure3}\textbf{b},
whereas the sensitivity of $^{87}$Rb magnetometer is only about $2~\textrm{pT} / \textrm{Hz}^{1/2}$.
Calibrated sensitivities with the spin-based amplifier from 2~Hz to 180~Hz are presented in Supplementary Information.
The experiment showcases the capability of our sensor to surpass the photon-shot-noise limit of the rubidium magnetometer itself, approaching the spin-projection-noise limit of the latter.
Moreover,
our sensing technique is significantly better than that of the other state-of-the-art magnetometers demonstrated with nuclear spins,
which are limited to a few picotesla sensitivity.

We perform the search for ALP signals in the frequency range from $2$~Hz to $180$~Hz,
corresponding to the ALP masses ranging from $8.3$~feV to $744$~feV.
In each run,
we set a fixed bias field, thus setting the ALP search frequency; and record 100~s of signal data.
We model the histogram of the power spectral values in the 2-Hz bin centered at $\nu_0$ as the chi-squared distribution with two degrees of freedom~\cite{walck2007hand},
and determine the standard deviation by fitting the cumulative distribution function.
In each run,
we can derive the $95\%$ confidence levels for $\mathrm{g}_{\textrm{aNN}}$ limits over the amplifier bandwidth, as shown in Fig.~\ref{figure4}\textbf{a}.
The data processing and exclusion procedure are discussed in the Supplementary Information.
By scanning bias field, limits on the axion-nucleon coupling constants $\textrm{g}_{\textrm{aNN}}$ over entire mass range are obtained as shown in Fig.~\ref{figure4}\textbf{b} (blue line).
The present work explores a mass range from $15$~feV to $78$~feV overlapping with the CASPEr-ZULF experiment~\cite{garcon2019constraints},
and improves the previous limits by more than five orders of magnitude.
Recent work~\cite{bloch2020axion} derives ALP limits below 200~feV based on the decade-old comagnetometer data from previously published work.
That work cannot access ALPs at higher Compton frequencies due to the sensitivity loss of comagnetometers~\cite{kornack2005nuclear}.
In contrast, we extend the ALP search to new unconstrained regions of the parameter space from 200~feV to 744~feV.
As a check of the viability of our data analysis procedure,
we inserted simulated ALP signals into our data and verified that the analysis method can recover the resonant and near-resonant ALP signals with their correct coupling strengths (Supplementary Information).
We note that preliminary theoretical investigations~\cite{centers2019stochastic} in a recent work show that if the amplitude fluctuations are taken into account for ultralight ALPs, then the inferred limits may be weakened by factors of order unity at $95\%$ confidence level.
However, these investigations are still at a preliminary stage and are beyond the scope of this article.

We note that the experimental sensitivity to $\textrm{g}_{\textrm{aNN}}$ scales as $t^{-1/2}$ as a function of measurement time $t$.
As an initial search, we report experimental results for 100 probed ALP-mass windows (see Fig.~\ref{figure4}\textbf{b}, red line) centered at different bias fields and with the width of 0.15~feV,
where each window is from five hours of data.
Comparing with 100-s search sensitivity,
we achieve one order of magnitude improvement on the ALP search sensitivity,
for example at 67.5 feV reaching $2.9 \times 10^{-9}$~GeV$^{-1}$ (95$\%$ confidence level).
We note that our laboratory limits are comparable to the stringent astrophysical limits obtained from supernova 1987A cooling~\cite{vysotsskii1978some, raffelt2008astrophysical}.
To complete the 5-h ALP searches in the entire range from 1~Hz to 200~Hz,
we propose to set up multiple spin-based amplifiers and cooperate with other research groups~\cite{groups} that have similar setups already in existence,
enabling to simultaneously search for ALPs and thus reduce the total acquisition time down to one month.
Moreover, the Global Network of Optical Magnetometers for Exotic physics (GNOME)~\cite{pospelov2013detecting, masia2020analysis} that already has a number of very similar devices could potentially be configured to run the present resonant searches.

Although demonstrated for axion-like dark matter searches,
our sensing technique can be immediately applied to search for a broad range of exotic fields and forces predicted by theories beyond the standard model.
Recent theoretical developments~\cite{pospelov2013detecting, graham2015experimental, graham2018spin} have predicted the existence of other bosons that could be constituents of the dark matter, for example, dark photons.
Here we show that our measurement constrains the nuclear spin interactions with dark photons through the coupling of the dark photon electric field to the dark EDM (dEDM)~\cite{wu2019search,garcon2019constraints} $\textrm{g}_{\textrm{dEDM}}$ (see Fig.~\ref{figure4}\textbf{c})
and the coupling of the dark photon magnetic field to the dark magnetic dipole moment (dMDM)~\cite{wu2019search,garcon2019constraints} $\textrm{g}_{\textrm{dMDM}}$ (see Fig.~S12 of Supplementary Information).
We also constrain the nuclear spin quadratic coupling with axions $\textrm{g}_{\textrm{quad}}$~\cite{wu2019search,garcon2019constraints} (Fig.~\ref{figure4}\textbf{d}).
Our achieved limits surpass the previous laboratory limits by several orders of magnitude~\cite{garcon2019constraints, wu2019search} and are beyond the astrophysical limits over a large part of the explored mass range.
The above limits for dark photon-nucleon interactions are from the 100-s data set and could be further improved based on our initial 5-h data set with one order of magnitude.
The details of $\textrm{g}_{\textrm{dEDM}}$, $\textrm{g}_{\textrm{quad}}$, and $\textrm{g}_{\textrm{dMDM}}$ are presented in Supplementary Information.
In addition,
numerous theories predict that ALPs can mediate new forces between objects~\cite{arvanitaki2014resonantly, aggarwal2020characterization, lee2018improved, ji2018new, kim2019experimental}.
The approach~\cite{arvanitaki2014resonantly, aggarwal2020characterization} is proposed based on resonant effect between nuclear spins and the effective field induced by axion-mediated force and could substantially improve upon current experimental limits set by astrophysics,
but has not yet been demonstrated experimentally.
Our work has already demonstrated a feasible route towards resonant amplification of the signal from exotic ALP fields and is also suited to searching for axion-mediated forces.
With the use of an optimized spin-based amplifier and a mass rotor, as recently demonstrated in refs.~\cite{arvanitaki2014resonantly, ji2018new, aggarwal2020characterization},
it allows extending the search for new forces with better sensitivity than previous works.



A further improvement of the experimental sensitivity to axion-like dark matter is anticipated.
Scaling the volume of the vapor cell up to $100$~cm$^3$ could improve the $^{87}$Rb magnetometer sensitivity by a factor of about $10$;
further, it is possible to achieve better sensor performance using multi-pass vapor cells~\cite{cai2020herriott}.
For an optimized search, it is necessary to optimize the amplification performance of spin-based amplifiers for ALP signals.
The most dramatic way would be using $^3$He-K systems,
because $^{3}$He spins have longer coherence time ($T_2$$\sim$$1000$~s) and larger gyromagnetic ratio than those of $^{129}$Xe ($T_2$$\sim$$20$~s in this work),
allowing one to achieve an amplification factor of $\eta\approx 10^4$ (Supplementary Information).
In addition, the $^3$He-K system have five orders smaller spin-destruction cross section than that of $^{129}$Xe-Rb systems,
and thus the K magnetometer can still achieve a femtotesla sensitivity, as demonstrated in ref.~\cite{kornack2005nuclear};
thus, the magnetic sensitivity based on $^3$He spin amplifier could probably reach $1~\textrm{aT} / \textrm{Hz}^{1/2}$ level within the amplifier bandwidth,
yielding experimental sensitivities of $|\textrm{g}_{\textrm{aNN}}|$$ \sim$$ 10^{-13}~\textrm{GeV}^{-1}$,
$|\textrm{g}_{\textrm{dEDM}}|$$ \sim$$ 10^{-17}~\textrm{GeV}^{-1}$,
$|\textrm{g}_{\textrm{dMDM}}|$$ \sim$$ 10^{-13}~\textrm{GeV}^{-1}$,
and $|\textrm{g}_{\textrm{quad}}|$$ \sim$$ 10^{-8}~\textrm{GeV}^{-1}$.
Our approach can be extended to a network of synchronized sensors:
the apparatus used in our experiment is small-scale and inexpensive.
Such a sensor network is promising to compose an exotic field telescope array for multi-messenger astronomy, as recently proposed~\cite{dailey2020quantum, pospelov2013detecting}, and enables an improvement of the sensitivity with the inverse square root of sensor number and allows to distinguish the exotic-physics signal from spurious noise.
Moreover correlating the readouts of many sensors in such a network could help address the stochastic fluctuations of bosonic dark matter~\cite{centers2019stochastic}.

In conclusion,
we have developed a new ALP search technique incorporating a spin-based amplifier,
which simultaneously allows to amplify the ALP signal with at least two orders of magnitude improvement as well as the readout of the amplified ALP signal.
Using such a sensor, our ALP search has significantly improved the constraint with at least five orders of magnitude improvement over previous results~\cite{garcon2019constraints, wu2019search} and placed the experimental bounds over an unconstrained parameter space,
opening a feasible route towards a laboratory-scale search sensitivity comparable to and even beyond the stringent astrophysical limits~\cite{vysotsskii1978some, raffelt2008astrophysical}.
This technique can be immediately applied to search for well-motivated bosonic exotic fields~\cite{wu2019search, garcon2019constraints} and axion mediated forces~\cite{arvanitaki2014resonantly, lee2018improved, ji2018new, kim2019experimental}.
Although for ALP searches, our work provides a conceptually novel sensing technique assisted with a spin-based amplifier,
which is promising for reaching the attotesla magnetic sensitivity level.

~\




~\

\noindent
\textbf{Acknowledgment}. This work of M.J., H.W.S., and X.H.P. was supported by National Key Research and Development Program of China (grant no. 2018YFA0306600), National Natural Science Foundation of China (grants nos. 11661161018, 11927811), Anhui Initiative in Quantum Information Technologies (grant no. AHY050000). The work of D.B. and A.G. was supported by the Cluster of Excellence PRISMA+ funded by the German Research Foundation (DFG) within the German Excellence Strategy (Project ID 39083149), by the European Research Council (ERC) under the European Union Horizon 2020 research and innovation program (project Dark-OST, grant agreement No 695405), by the DFG Reinhart Koselleck project, and by the Emergent AI Center funded by the Carl-Zeiss-Stiftung.

~\

\noindent
\textbf{Author contributions}.
M.J. designed experimental protocols, analyzed the data and wrote the manuscript.
H.W.S. performed experiments, analyzed the data and wrote the manuscript.
A.G. analyzed the data and edited the manuscript.
X.H.P. proposed the experimental concept, devised the experimental protocols, and edited the manuscript.
D.B. contributed to the design of the experiment and proofread and edited the manuscript.
All authors contributed with discussions and checking the manuscript.

~\

\noindent
\textbf{Competing interests}.
The authors declare that they have no competing interests.

\bibliographystyle{naturemag}
\bibliography{mainrefs}

\begin{thebibliography}{10}
\expandafter\ifx\csname url\endcsname\relax
  \def\url#1{\texttt{#1}}\fi
\expandafter\ifx\csname urlprefix\endcsname\relax\def\urlprefix{URL }\fi
\providecommand{\bibinfo}[2]{#2}
\providecommand{\eprint}[2][]{\url{#2}}

\bibitem{bertone2018history}
\bibinfo{author}{Bertone, G.} \& \bibinfo{author}{Hooper, D.}
\newblock \bibinfo{title}{History of dark matter}.
\newblock \emph{\bibinfo{journal}{Rev. Mod. Phys.}}
  \textbf{\bibinfo{volume}{90}}, \bibinfo{pages}{045002}
  (\bibinfo{year}{2018}).

\bibitem{demille2017probing}
\bibinfo{author}{DeMille, D.}, \bibinfo{author}{Doyle, J.~M.} \&
  \bibinfo{author}{Sushkov, A.~O.}
\newblock \bibinfo{title}{Probing the frontiers of particle physics with
  tabletop-scale experiments}.
\newblock \emph{\bibinfo{journal}{Science}} \textbf{\bibinfo{volume}{357}},
  \bibinfo{pages}{990--994} (\bibinfo{year}{2017}).

\bibitem{safronova2018search}
\bibinfo{author}{Safronova, M.} \emph{et~al.}
\newblock \bibinfo{title}{Search for new physics with atoms and molecules}.
\newblock \emph{\bibinfo{journal}{Rev. Mod. Phys.}}
  \textbf{\bibinfo{volume}{90}}, \bibinfo{pages}{025008}
  (\bibinfo{year}{2018}).

\bibitem{bertone2018new}
\bibinfo{author}{Bertone, G.} \& \bibinfo{author}{Tait, T.~M.}
\newblock \bibinfo{title}{A new era in the search for dark matter}.
\newblock \emph{\bibinfo{journal}{Nature}} \textbf{\bibinfo{volume}{562}},
  \bibinfo{pages}{51--56} (\bibinfo{year}{2018}).

\bibitem{ambrosi2017direct}
\bibinfo{author}{Ambrosi, G.} \emph{et~al.}
\newblock \bibinfo{title}{Direct detection of a break in the teraelectronvolt
  cosmic-ray spectrum of electrons and positrons}.
\newblock \emph{\bibinfo{journal}{Nature}} \textbf{\bibinfo{volume}{552}},
  \bibinfo{pages}{63--66} (\bibinfo{year}{2017}).

\bibitem{aprile2017first}
\bibinfo{author}{Aprile, E.} \emph{et~al.}
\newblock \bibinfo{title}{First dark matter search results from the xenon1t
  experiment}.
\newblock \emph{\bibinfo{journal}{Phys. Rev. Lett.}}
  \textbf{\bibinfo{volume}{119}}, \bibinfo{pages}{181301}
  (\bibinfo{year}{2017}).

\bibitem{liu2017current}
\bibinfo{author}{Liu, J.}, \bibinfo{author}{Chen, X.} \& \bibinfo{author}{Ji,
  X.}
\newblock \bibinfo{title}{Current status of direct dark matter detection
  experiments}.
\newblock \emph{\bibinfo{journal}{Nat. Phys.}} \textbf{\bibinfo{volume}{13}},
  \bibinfo{pages}{212--216} (\bibinfo{year}{2017}).

\bibitem{peccei1977cp}
\bibinfo{author}{Peccei, R.~D.} \& \bibinfo{author}{Quinn, H.~R.}
\newblock \bibinfo{title}{\rm{CP} conservation in the presence of
  pseudoparticles}.
\newblock \emph{\bibinfo{journal}{Phys. Rev. Lett.}}
  \textbf{\bibinfo{volume}{38}}, \bibinfo{pages}{1440} (\bibinfo{year}{1977}).

\bibitem{peccei1977constraints}
\bibinfo{author}{Peccei, R.~D.} \& \bibinfo{author}{Quinn, H.~R.}
\newblock \bibinfo{title}{Constraints imposed by \rm{CP} conservation in the
  presence of pseudoparticles}.
\newblock \emph{\bibinfo{journal}{Phys. Rev. D}} \textbf{\bibinfo{volume}{16}},
  \bibinfo{pages}{1791} (\bibinfo{year}{1977}).

\bibitem{preskill1983cosmology}
\bibinfo{author}{Preskill, J.}, \bibinfo{author}{Wise, M.~B.} \&
  \bibinfo{author}{Wilczek, F.}
\newblock \bibinfo{title}{Cosmology of the invisible axion}.
\newblock \emph{\bibinfo{journal}{Phys. Lett. B}}
  \textbf{\bibinfo{volume}{120}}, \bibinfo{pages}{127--132}
  (\bibinfo{year}{1983}).

\bibitem{kim2010axions}
\bibinfo{author}{Kim, J.~E.} \& \bibinfo{author}{Carosi, G.}
\newblock \bibinfo{title}{Axions and the strong \rm{CP} problem}.
\newblock \emph{\bibinfo{journal}{Rev. Mod. Phys.}}
  \textbf{\bibinfo{volume}{82}}, \bibinfo{pages}{557} (\bibinfo{year}{2010}).

\bibitem{irastorza2018new}
\bibinfo{author}{Irastorza, I.~G.} \& \bibinfo{author}{Redondo, J.}
\newblock \bibinfo{title}{New experimental approaches in the search for
  axion-like particles}.
\newblock \emph{\bibinfo{journal}{Prog. Part. Nucl. Phys.}}
  \textbf{\bibinfo{volume}{102}}, \bibinfo{pages}{89--159}
  (\bibinfo{year}{2018}).

\bibitem{svrcek2006axions}
\bibinfo{author}{Svrcek, P.} \& \bibinfo{author}{Witten, E.}
\newblock \bibinfo{title}{Axions in string theory}.
\newblock \emph{\bibinfo{journal}{J. High Energy Phys.}}
  \textbf{\bibinfo{volume}{2006}}, \bibinfo{pages}{051} (\bibinfo{year}{2006}).

\bibitem{anastassopoulos2017new}
\bibinfo{author}{Anastassopoulos, V.} \emph{et~al.}
\newblock \bibinfo{title}{New \rm{CAST} limit on the axion--photon
  interaction}.
\newblock \emph{\bibinfo{journal}{Nat. Phys.}} \textbf{\bibinfo{volume}{13}},
  \bibinfo{pages}{584} (\bibinfo{year}{2017}).

\bibitem{bradley2003microwave}
\bibinfo{author}{Bradley, R.} \emph{et~al.}
\newblock \bibinfo{title}{Microwave cavity searches for dark-matter axions}.
\newblock \emph{\bibinfo{journal}{Rev. Mod. Phys.}}
  \textbf{\bibinfo{volume}{75}}, \bibinfo{pages}{777} (\bibinfo{year}{2003}).

\bibitem{zhong2018results}
\bibinfo{author}{Zhong, L.} \emph{et~al.}
\newblock \bibinfo{title}{Results from phase 1 of the \rm{HAYSTAC} microwave
  cavity axion experiment}.
\newblock \emph{\bibinfo{journal}{Phys. Rev. D}} \textbf{\bibinfo{volume}{97}},
  \bibinfo{pages}{092001} (\bibinfo{year}{2018}).

\bibitem{braine2020extended}
\bibinfo{author}{Braine, T.} \emph{et~al.}
\newblock \bibinfo{title}{Extended search for the invisible axion with the
  axion dark matter experiment}.
\newblock \emph{\bibinfo{journal}{Phys. Rev. Lett.}}
  \textbf{\bibinfo{volume}{124}}, \bibinfo{pages}{101303}
  (\bibinfo{year}{2020}).

\bibitem{ouellet2019first}
\bibinfo{author}{Ouellet, J.~L.} \emph{et~al.}
\newblock \bibinfo{title}{First results from \rm{ABRACADABRA}-10 cm: A search
  for sub-$\mu$ev axion dark matter}.
\newblock \emph{\bibinfo{journal}{Phys. Rev. Lett.}}
  \textbf{\bibinfo{volume}{122}}, \bibinfo{pages}{121802}
  (\bibinfo{year}{2019}).

\bibitem{gramolin2020search}
\bibinfo{author}{Gramolin, A.~V.}, \bibinfo{author}{Aybas, D.},
  \bibinfo{author}{Johnson, D.}, \bibinfo{author}{Adam, J.} \&
  \bibinfo{author}{Sushkov, A.~O.}
\newblock \bibinfo{title}{Search for axion-like dark matter with ferromagnets}.
\newblock \emph{\bibinfo{journal}{Nat. Phys.}} \textbf{\bibinfo{volume}{17}},
  \bibinfo{pages}{79--84} (\bibinfo{year}{2020}).

\bibitem{budker2014proposal}
\bibinfo{author}{Budker, D.}, \bibinfo{author}{Graham, P.~W.},
  \bibinfo{author}{Ledbetter, M.}, \bibinfo{author}{Rajendran, S.} \&
  \bibinfo{author}{Sushkov, A.~O.}
\newblock \bibinfo{title}{Proposal for a cosmic axion spin precession
  experiment (\rm{CASPEr})}.
\newblock \emph{\bibinfo{journal}{Phys. Rev. X}} \textbf{\bibinfo{volume}{4}},
  \bibinfo{pages}{021030} (\bibinfo{year}{2014}).

\bibitem{roberts2014limiting}
\bibinfo{author}{Roberts, B.} \emph{et~al.}
\newblock \bibinfo{title}{Limiting \rm{P}-odd interactions of cosmic fields
  with electrons, protons, and neutrons}.
\newblock \emph{\bibinfo{journal}{Phys. Rev. Lett.}}
  \textbf{\bibinfo{volume}{113}}, \bibinfo{pages}{081601}
  (\bibinfo{year}{2014}).

\bibitem{graham2011axion}
\bibinfo{author}{Graham, P.~W.} \& \bibinfo{author}{Rajendran, S.}
\newblock \bibinfo{title}{Axion dark matter detection with cold molecules}.
\newblock \emph{\bibinfo{journal}{Phys. Rev. D}} \textbf{\bibinfo{volume}{84}},
  \bibinfo{pages}{055013} (\bibinfo{year}{2011}).

\bibitem{stadnik2014axion}
\bibinfo{author}{Stadnik, Y.} \& \bibinfo{author}{Flambaum, V.}
\newblock \bibinfo{title}{Axion-induced effects in atoms, molecules, and
  nuclei: \rm{Parity} nonconservation, anapole moments, electric dipole
  moments, and spin-gravity and spin-axion momentum couplings}.
\newblock \emph{\bibinfo{journal}{Phys. Rev. D}} \textbf{\bibinfo{volume}{89}},
  \bibinfo{pages}{043522} (\bibinfo{year}{2014}).

\bibitem{kimball2020overview}
\bibinfo{author}{Kimball, D. F.~J.} \emph{et~al.}
\newblock \bibinfo{title}{Overview of the cosmic axion spin precession
  experiment (\rm{CASPEr})}.
\newblock In \emph{\bibinfo{booktitle}{Microwave Cavities and Detectors for
  Axion Research}}, \bibinfo{pages}{105--121} (\bibinfo{publisher}{Springer},
  \bibinfo{year}{2020}).

\bibitem{jiang2019floquet}
\bibinfo{author}{Jiang, M.}, \bibinfo{author}{Su, H.}, \bibinfo{author}{Wu,
  Z.}, \bibinfo{author}{Peng, X.} \& \bibinfo{author}{Budker, D.}
\newblock \bibinfo{title}{Floquet maser}.
\newblock \emph{\bibinfo{journal}{\rm{Preprint at
  https://arxiv.org/abs/1901.00970 (2019)}; $Sci.~ Adv.$}}
  \textbf{\bibinfo{volume}{7}}, \bibinfo{pages}{eabe0719}
  (\bibinfo{year}{2021}).

\bibitem{abel2017search}
\bibinfo{author}{Abel, C.} \emph{et~al.}
\newblock \bibinfo{title}{Search for axionlike dark matter through nuclear spin
  precession in electric and magnetic fields}.
\newblock \emph{\bibinfo{journal}{Phys. Rev. X}} \textbf{\bibinfo{volume}{7}},
  \bibinfo{pages}{041034} (\bibinfo{year}{2017}).

\bibitem{wu2019search}
\bibinfo{author}{Wu, T.} \emph{et~al.}
\newblock \bibinfo{title}{Search for axionlike dark matter with a liquid-state
  nuclear spin comagnetometer}.
\newblock \emph{\bibinfo{journal}{Phys. Rev. Lett.}}
  \textbf{\bibinfo{volume}{122}}, \bibinfo{pages}{191302}
  (\bibinfo{year}{2019}).

\bibitem{smorra2019direct}
\bibinfo{author}{Smorra, C.} \emph{et~al.}
\newblock \bibinfo{title}{Direct limits on the interaction of antiprotons with
  axion-like dark matter}.
\newblock \emph{\bibinfo{journal}{Nature}} \textbf{\bibinfo{volume}{575}},
  \bibinfo{pages}{310--314} (\bibinfo{year}{2019}).

\bibitem{garcon2019constraints}
\bibinfo{author}{Garcon, A.} \emph{et~al.}
\newblock \bibinfo{title}{Constraints on bosonic dark matter from
  ultralow-field nuclear magnetic resonance}.
\newblock \emph{\bibinfo{journal}{Sci. Adv.}} \textbf{\bibinfo{volume}{5}},
  \bibinfo{pages}{eaax4539} (\bibinfo{year}{2019}).

\bibitem{graham2018spin}
\bibinfo{author}{Graham, P.~W.} \emph{et~al.}
\newblock \bibinfo{title}{Spin precession experiments for light axionic dark
  matter}.
\newblock \emph{\bibinfo{journal}{Phys. Rev. D}} \textbf{\bibinfo{volume}{97}},
  \bibinfo{pages}{055006} (\bibinfo{year}{2018}).

\bibitem{graham2013new}
\bibinfo{author}{Graham, P.~W.} \& \bibinfo{author}{Rajendran, S.}
\newblock \bibinfo{title}{New observables for direct detection of axion dark
  matter}.
\newblock \emph{\bibinfo{journal}{Phys. Rev. D}} \textbf{\bibinfo{volume}{88}},
  \bibinfo{pages}{035023} (\bibinfo{year}{2013}).

\bibitem{aybas2021search}
\bibinfo{author}{Aybas, D.} \emph{et~al.}
\newblock \bibinfo{title}{Search for axion-like dark matter using solid-state
  nuclear magnetic resonance.} \bibinfo{pages}{Preprint at
  https://arxiv.org/abs/2101.01241} (\bibinfo{year}{2021}).

\bibitem{bloch2020axion}
\bibinfo{author}{Bloch, I.~M.}, \bibinfo{author}{Hochberg, Y.},
  \bibinfo{author}{Kuflik, E.} \& \bibinfo{author}{Volansky, T.}
\newblock \bibinfo{title}{Axion-like relics: new constraints from old
  comagnetometer data}.
\newblock \emph{\bibinfo{journal}{J. High Energy Phys.}}
  \textbf{\bibinfo{volume}{2020}}, \bibinfo{pages}{1--38}
  (\bibinfo{year}{2020}).

\bibitem{vysotsskii1978some}
\bibinfo{author}{Vysotsskii, M.}, \bibinfo{author}{Zel’Dovich, Y.~B.},
  \bibinfo{author}{Khlopov, M.~Y.} \& \bibinfo{author}{Chechetkin, V.}
\newblock \bibinfo{title}{Some astrophysical limitations on the axion mass}.
\newblock \emph{\bibinfo{journal}{JETP Lett.}} \textbf{\bibinfo{volume}{27}},
  \bibinfo{pages}{533} (\bibinfo{year}{1978}).

\bibitem{raffelt2008astrophysical}
\bibinfo{author}{Raffelt, G.~G.}
\newblock \bibinfo{title}{Astrophysical axion bounds}.
\newblock In \emph{\bibinfo{booktitle}{Axions}}, \bibinfo{pages}{51--71}
  (\bibinfo{publisher}{Springer}, \bibinfo{year}{2008}).

\bibitem{kominis2003subfemtotesla}
\bibinfo{author}{Kominis, I.}, \bibinfo{author}{Kornack, T.},
  \bibinfo{author}{Allred, J.} \& \bibinfo{author}{Romalis, M.~V.}
\newblock \bibinfo{title}{A subfemtotesla multichannel atomic magnetometer}.
\newblock \emph{\bibinfo{journal}{Nature}} \textbf{\bibinfo{volume}{422}},
  \bibinfo{pages}{596--599} (\bibinfo{year}{2003}).

\bibitem{budker2007optical}
\bibinfo{author}{Budker, D.} \& \bibinfo{author}{Romalis, M.}
\newblock \bibinfo{title}{Optical magnetometry}.
\newblock \emph{\bibinfo{journal}{Nat. Phys.}} \textbf{\bibinfo{volume}{3}},
  \bibinfo{pages}{227--234} (\bibinfo{year}{2007}).

\bibitem{arvanitaki2014resonantly}
\bibinfo{author}{Arvanitaki, A.} \& \bibinfo{author}{Geraci, A.~A.}
\newblock \bibinfo{title}{Resonantly detecting axion-mediated forces with
  nuclear magnetic resonance}.
\newblock \emph{\bibinfo{journal}{Phys. Rev. Lett.}}
  \textbf{\bibinfo{volume}{113}}, \bibinfo{pages}{161801}
  (\bibinfo{year}{2014}).

\bibitem{walker1997spin}
\bibinfo{author}{Walker, T.~G.} \& \bibinfo{author}{Happer, W.}
\newblock \bibinfo{title}{Spin-exchange optical pumping of noble-gas nuclei}.
\newblock \emph{\bibinfo{journal}{Rev. Mod. Phys.}}
  \textbf{\bibinfo{volume}{69}}, \bibinfo{pages}{629} (\bibinfo{year}{1997}).

\bibitem{marsh2016axion}
\bibinfo{author}{Marsh, D.~J.}
\newblock \bibinfo{title}{Axion cosmology}.
\newblock \emph{\bibinfo{journal}{Phys. Rep.}} \textbf{\bibinfo{volume}{643}},
  \bibinfo{pages}{1--79} (\bibinfo{year}{2016}).

\bibitem{dine1983not}
\bibinfo{author}{Dine, M.} \& \bibinfo{author}{Fischler, W.}
\newblock \bibinfo{title}{The not-so-harmless axion}.
\newblock \emph{\bibinfo{journal}{Phys. Lett. B}}
  \textbf{\bibinfo{volume}{120}}, \bibinfo{pages}{137--141}
  (\bibinfo{year}{1983}).

\bibitem{walck2007hand}
\bibinfo{author}{Walck, C.}
\newblock \bibinfo{title}{Hand-book on statistical distributions for
  experimentalists.} \bibinfo{pages}{Internal Report SUF--PFY/96--01}
  (\bibinfo{year}{Univ. Stockholm, 2007}).

\bibitem{kornack2005nuclear}
\bibinfo{author}{Kornack, T.}, \bibinfo{author}{Ghosh, R.} \&
  \bibinfo{author}{Romalis, M.}
\newblock \bibinfo{title}{Nuclear spin gyroscope based on an atomic
  comagnetometer}.
\newblock \emph{\bibinfo{journal}{Phys. Rev. Lett.}}
  \textbf{\bibinfo{volume}{95}}, \bibinfo{pages}{230801}
  (\bibinfo{year}{2005}).

\bibitem{centers2019stochastic}
\bibinfo{author}{Centers, G.~P.} \emph{et~al.}
\newblock \bibinfo{title}{Stochastic fluctuations of bosonic dark matter.}
  \bibinfo{pages}{Preprint at https://arxiv.org/abs/1905.13650}
  (\bibinfo{year}{2019}).

\bibitem{groups}
\emph{\bibinfo{journal}{\rm{University of Science and Technology of China;
  Johannes Gutenberg University; Peking University; Tsinghua University;
  Zhejiang University Of Technology; Wuhan Institute of Physics and Mathematics
  (WIPM) of Chinese Academy of Sciences; Beijing Computational Science Research
  Center; China Shipbuilding Industry Corporation 707 Research Institute;
  Suzhou Institute of Nano-Tech and Nano-Bionics}}} .

\bibitem{pospelov2013detecting}
\bibinfo{author}{Pospelov, M.} \emph{et~al.}
\newblock \bibinfo{title}{Detecting domain walls of axionlike models using
  terrestrial experiments}.
\newblock \emph{\bibinfo{journal}{Phys. Rev. Lett.}}
  \textbf{\bibinfo{volume}{110}}, \bibinfo{pages}{021803}
  (\bibinfo{year}{2013}).

\bibitem{masia2020analysis}
\bibinfo{author}{Masia-Roig, H.} \emph{et~al.}
\newblock \bibinfo{title}{Analysis method for detecting topological defect dark
  matter with a global magnetometer network}.
\newblock \emph{\bibinfo{journal}{Phys. Dark Universe}}
  \textbf{\bibinfo{volume}{28}}, \bibinfo{pages}{100494}
  (\bibinfo{year}{2020}).

\bibitem{graham2015experimental}
\bibinfo{author}{Graham, P.~W.}, \bibinfo{author}{Irastorza, I.~G.},
  \bibinfo{author}{Lamoreaux, S.~K.}, \bibinfo{author}{Lindner, A.} \&
  \bibinfo{author}{van Bibber, K.~A.}
\newblock \bibinfo{title}{Experimental searches for the axion and axion-like
  particles}.
\newblock \emph{\bibinfo{journal}{Annu. Rev. Nucl. Part. Sci.}}
  \textbf{\bibinfo{volume}{65}}, \bibinfo{pages}{485--514}
  (\bibinfo{year}{2015}).

\bibitem{aggarwal2020characterization}
\bibinfo{author}{Aggarwal, N.} \emph{et~al.}
\newblock \bibinfo{title}{Characterization of magnetic field noise in the
  \rm{ARIADNE} source mass rotor.} \bibinfo{pages}{Preprint at
  https://arxiv.org/abs/2011.12617} (\bibinfo{year}{2020}).

\bibitem{lee2018improved}
\bibinfo{author}{Lee, J.}, \bibinfo{author}{Almasi, A.} \&
  \bibinfo{author}{Romalis, M.}
\newblock \bibinfo{title}{Improved limits on spin-mass interactions}.
\newblock \emph{\bibinfo{journal}{Phys. Rev. Lett.}}
  \textbf{\bibinfo{volume}{120}}, \bibinfo{pages}{161801}
  (\bibinfo{year}{2018}).

\bibitem{ji2018new}
\bibinfo{author}{Ji, W.} \emph{et~al.}
\newblock \bibinfo{title}{\rm{New Experimental Limits on Exotic
  Spin-Spin-Velocity-Dependent Interactions by Using SmCo5 Spin Sources}}.
\newblock \emph{\bibinfo{journal}{Phys. Rev. Lett.}}
  \textbf{\bibinfo{volume}{121}}, \bibinfo{pages}{261803}
  (\bibinfo{year}{2018}).

\bibitem{kim2019experimental}
\bibinfo{author}{Kim, Y.~J.}, \bibinfo{author}{Chu, P.-H.},
  \bibinfo{author}{Savukov, I.} \& \bibinfo{author}{Newman, S.}
\newblock \bibinfo{title}{Experimental limit on an exotic parity-odd spin-and
  velocity-dependent interaction using an optically polarized vapor}.
\newblock \emph{\bibinfo{journal}{Nat. Commun.}} \textbf{\bibinfo{volume}{10}},
  \bibinfo{pages}{1--7} (\bibinfo{year}{2019}).

\bibitem{cai2020herriott}
\bibinfo{author}{Cai, B.} \emph{et~al.}
\newblock \bibinfo{title}{Herriott-cavity-assisted all-optical atomic vector
  magnetometer}.
\newblock \emph{\bibinfo{journal}{Phys. Rev. A}}
  \textbf{\bibinfo{volume}{101}}, \bibinfo{pages}{053436}
  (\bibinfo{year}{2020}).

\bibitem{dailey2020quantum}
\bibinfo{author}{Dailey, C.} \emph{et~al.}
\newblock \bibinfo{title}{Quantum sensor networks as exotic field telescopes
  for multi-messenger astronomy}.
\newblock \emph{\bibinfo{journal}{Nat. Astron.,}} \bibinfo{pages}{1--9}
  (\bibinfo{year}{2020}).

\end{thebibliography}


\begin{thebibliography}{10}
\expandafter\ifx\csname url\endcsname\relax
  \def\url#1{\texttt{#1}}\fi
\expandafter\ifx\csname urlprefix\endcsname\relax\def\urlprefix{URL }\fi
\providecommand{\bibinfo}[2]{#2}
\providecommand{\eprint}[2][]{\url{#2}}

\bibitem{wu1986optical}
\bibinfo{author}{Wu, Z.}, \bibinfo{author}{Kitano, M.},
  \bibinfo{author}{Happer, W.}, \bibinfo{author}{Hou, M.} \&
  \bibinfo{author}{Daniels, J.}
\newblock \bibinfo{title}{Optical determination of alkali metal vapor number
  density using faraday rotation}.
\newblock \emph{\bibinfo{journal}{Appl. Opt.}} \textbf{\bibinfo{volume}{25}},
  \bibinfo{pages}{4483--4492} (\bibinfo{year}{1986}).

\bibitem{opechowski1953magneto}
\bibinfo{author}{Opechowski, W.}
\newblock \bibinfo{title}{Magneto-optical effects and paramagnetic resonance}.
\newblock \emph{\bibinfo{journal}{Rev. Mod. Phys.}}
  \textbf{\bibinfo{volume}{25}}, \bibinfo{pages}{264} (\bibinfo{year}{1953}).

\bibitem{jiang2020interference}
\bibinfo{author}{Jiang, M.} \emph{et~al.}
\newblock \bibinfo{title}{Interference in atomic magnetometry}.
\newblock \emph{\bibinfo{journal}{Adv. Quantum Technol.}}
  \textbf{\bibinfo{volume}{3}}, \bibinfo{pages}{2000078}
  (\bibinfo{year}{2020}).

\bibitem{kornack2005nuclear}
\bibinfo{author}{Kornack, T.}, \bibinfo{author}{Ghosh, R.} \&
  \bibinfo{author}{Romalis, M.}
\newblock \bibinfo{title}{Nuclear spin gyroscope based on an atomic
  comagnetometer}.
\newblock \emph{\bibinfo{journal}{Phys. Rev. Lett.}}
  \textbf{\bibinfo{volume}{95}}, \bibinfo{pages}{230801}
  (\bibinfo{year}{2005}).

\bibitem{walker1997spin}
\bibinfo{author}{Walker, T.~G.} \& \bibinfo{author}{Happer, W.}
\newblock \bibinfo{title}{Spin-exchange optical pumping of noble-gas nuclei}.
\newblock \emph{\bibinfo{journal}{Rev. Mod. Phys.}}
  \textbf{\bibinfo{volume}{69}}, \bibinfo{pages}{629} (\bibinfo{year}{1997}).

\bibitem{kimball2020overview}
\bibinfo{author}{Kimball, D. F.~J.} \emph{et~al.}
\newblock \bibinfo{title}{Overview of the cosmic axion spin precession
  experiment (\rm{CASPEr})}.
\newblock In \emph{\bibinfo{booktitle}{Microwave Cavities and Detectors for
  Axion Research}}, \bibinfo{pages}{105--121} (\bibinfo{publisher}{Springer},
  \bibinfo{year}{2020}).

\bibitem{abel2017search}
\bibinfo{author}{Abel, C.} \emph{et~al.}
\newblock \bibinfo{title}{Search for axionlike dark matter through nuclear spin
  precession in electric and magnetic fields}.
\newblock \emph{\bibinfo{journal}{Phys. Rev. X}} \textbf{\bibinfo{volume}{7}},
  \bibinfo{pages}{041034} (\bibinfo{year}{2017}).

\bibitem{wu2019search}
\bibinfo{author}{Wu, T.} \emph{et~al.}
\newblock \bibinfo{title}{Search for axionlike dark matter with a liquid-state
  nuclear spin comagnetometer}.
\newblock \emph{\bibinfo{journal}{Phys. Rev. Lett.}}
  \textbf{\bibinfo{volume}{122}}, \bibinfo{pages}{191302}
  (\bibinfo{year}{2019}).

\bibitem{smorra2019direct}
\bibinfo{author}{Smorra, C.} \emph{et~al.}
\newblock \bibinfo{title}{Direct limits on the interaction of antiprotons with
  axion-like dark matter}.
\newblock \emph{\bibinfo{journal}{Nature}} \textbf{\bibinfo{volume}{575}},
  \bibinfo{pages}{310--314} (\bibinfo{year}{2019}).

\bibitem{garcon2019constraints}
\bibinfo{author}{Garcon, A.} \emph{et~al.}
\newblock \bibinfo{title}{Constraints on bosonic dark matter from
  ultralow-field nuclear magnetic resonance}.
\newblock \emph{\bibinfo{journal}{Sci. Adv.}} \textbf{\bibinfo{volume}{5}},
  \bibinfo{pages}{eaax4539} (\bibinfo{year}{2019}).

\bibitem{graham2018spin}
\bibinfo{author}{Graham, P.~W.} \emph{et~al.}
\newblock \bibinfo{title}{Spin precession experiments for light axionic dark
  matter}.
\newblock \emph{\bibinfo{journal}{Phys. Rev. D}} \textbf{\bibinfo{volume}{97}},
  \bibinfo{pages}{055006} (\bibinfo{year}{2018}).

\bibitem{graham2013new}
\bibinfo{author}{Graham, P.~W.} \& \bibinfo{author}{Rajendran, S.}
\newblock \bibinfo{title}{New observables for direct detection of axion dark
  matter}.
\newblock \emph{\bibinfo{journal}{Phys. Rev. D}} \textbf{\bibinfo{volume}{88}},
  \bibinfo{pages}{035023} (\bibinfo{year}{2013}).

\bibitem{marsh2016axion}
\bibinfo{author}{Marsh, D.~J.}
\newblock \bibinfo{title}{Axion cosmology}.
\newblock \emph{\bibinfo{journal}{Phys. Rep.}} \textbf{\bibinfo{volume}{643}},
  \bibinfo{pages}{1--79} (\bibinfo{year}{2016}).

\bibitem{kostelecky1999constraints}
\bibinfo{author}{Kosteleck{\`y}, V.~A.} \& \bibinfo{author}{Lane, C.~D.}
\newblock \bibinfo{title}{Constraints on lorentz violation from
  clock-comparison experiments}.
\newblock \emph{\bibinfo{journal}{Phys. Rev. D}} \textbf{\bibinfo{volume}{60}},
  \bibinfo{pages}{116010} (\bibinfo{year}{1999}).

\bibitem{walck2007hand}
\bibinfo{author}{Walck, C.}
\newblock \bibinfo{title}{Hand-book on statistical distributions for
  experimentalists.} \bibinfo{pages}{Internal Report SUF--PFY/96--01}
  (\bibinfo{year}{Univ. Stockholm, 2007}).

\bibitem{pospelov2013detecting}
\bibinfo{author}{Pospelov, M.} \emph{et~al.}
\newblock \bibinfo{title}{Detecting domain walls of axionlike models using
  terrestrial experiments}.
\newblock \emph{\bibinfo{journal}{Phys. Rev. Lett.}}
  \textbf{\bibinfo{volume}{110}}, \bibinfo{pages}{021803}
  (\bibinfo{year}{2013}).

\bibitem{graham2015experimental}
\bibinfo{author}{Graham, P.~W.}, \bibinfo{author}{Irastorza, I.~G.},
  \bibinfo{author}{Lamoreaux, S.~K.}, \bibinfo{author}{Lindner, A.} \&
  \bibinfo{author}{van Bibber, K.~A.}
\newblock \bibinfo{title}{Experimental searches for the axion and axion-like
  particles}.
\newblock \emph{\bibinfo{journal}{Annu. Rev. Nucl. Part. Sci.}}
  \textbf{\bibinfo{volume}{65}}, \bibinfo{pages}{485--514}
  (\bibinfo{year}{2015}).

\bibitem{vysotsskii1978some}
\bibinfo{author}{Vysotsskii, M.}, \bibinfo{author}{Zel’Dovich, Y.~B.},
  \bibinfo{author}{Khlopov, M.~Y.} \& \bibinfo{author}{Chechetkin, V.}
\newblock \bibinfo{title}{Some astrophysical limitations on the axion mass}.
\newblock \emph{\bibinfo{journal}{JETP Lett.}} \textbf{\bibinfo{volume}{27}},
  \bibinfo{pages}{533} (\bibinfo{year}{1978}).

\bibitem{raffelt2008astrophysical}
\bibinfo{author}{Raffelt, G.~G.}
\newblock \bibinfo{title}{Astrophysical axion bounds}.
\newblock In \emph{\bibinfo{booktitle}{Axions}}, \bibinfo{pages}{51--71}
  (\bibinfo{publisher}{Springer}, \bibinfo{year}{2008}).

\bibitem{kominis2003subfemtotesla}
\bibinfo{author}{Kominis, I.}, \bibinfo{author}{Kornack, T.},
  \bibinfo{author}{Allred, J.} \& \bibinfo{author}{Romalis, M.~V.}
\newblock \bibinfo{title}{A subfemtotesla multichannel atomic magnetometer}.
\newblock \emph{\bibinfo{journal}{Nature}} \textbf{\bibinfo{volume}{422}},
  \bibinfo{pages}{596--599} (\bibinfo{year}{2003}).

\bibitem{dailey2020quantum}
\bibinfo{author}{Dailey, C.} \emph{et~al.}
\newblock \bibinfo{title}{Quantum sensor networks as exotic field telescopes
  for multi-messenger astronomy}.
\newblock \emph{\bibinfo{journal}{Nat. Astron.,}} \bibinfo{pages}{1--9}
  (\bibinfo{year}{2020}).

\bibitem{centers2019stochastic}
\bibinfo{author}{Centers, G.~P.} \emph{et~al.}
\newblock \bibinfo{title}{Stochastic fluctuations of bosonic dark matter.}
  \bibinfo{pages}{Preprint at https://arxiv.org/abs/1905.13650}
  (\bibinfo{year}{2019}).

\end{thebibliography}

\end{document}


\title{Supplementary Information for: \\``Search for axion-like dark matter with spin-based amplifiers"}

\date{\today}

\author{Min Jiang}
\email[]{These authors contributed equally to this work}
\affiliation{
Hefei National Laboratory for Physical Sciences at the Microscale and Department of Modern Physics, University of Science and Technology of China, Hefei 230026, China}
\affiliation{
CAS Key Laboratory of Microscale Magnetic Resonance, University of Science and Technology of China, Hefei 230026, China}
\affiliation{
Synergetic Innovation Center of Quantum Information and Quantum Physics, University of Science and Technology of China, Hefei 230026, China}

\author{Haowen Su}
\email[]{These authors contributed equally to this work}
\affiliation{
Hefei National Laboratory for Physical Sciences at the Microscale and Department of Modern Physics, University of Science and Technology of China, Hefei 230026, China}
\affiliation{
CAS Key Laboratory of Microscale Magnetic Resonance, University of Science and Technology of China, Hefei 230026, China}
\affiliation{
Synergetic Innovation Center of Quantum Information and Quantum Physics, University of Science and Technology of China, Hefei 230026, China}

\author{Antoine Garcon}
\affiliation{Helmholtz-Institut, GSI Helmholtzzentrum f{\"u}r Schwerionenforschung, Mainz 55128, Germany}
\affiliation{Johannes Gutenberg University, Mainz 55128, Germany}

\author{Xinhua Peng}
\email[]{xhpeng@ustc.edu.cn}
\affiliation{
Hefei National Laboratory for Physical Sciences at the Microscale and Department of Modern Physics, University of Science and Technology of China, Hefei 230026, China}
\affiliation{
CAS Key Laboratory of Microscale Magnetic Resonance, University of Science and Technology of China, Hefei 230026, China}
\affiliation{
Synergetic Innovation Center of Quantum Information and Quantum Physics, University of Science and Technology of China, Hefei 230026, China}

\author{Dmitry Budker}
\affiliation{Helmholtz-Institut, GSI Helmholtzzentrum f{\"u}r Schwerionenforschung, Mainz 55128, Germany}
\affiliation{Johannes Gutenberg University, Mainz 55128, Germany}
\affiliation{Department of Physics, University of California, Berkeley, CA 94720-7300, USA}

\maketitle

\tableofcontents

\begin{figure*}[t]  
	\makeatletter
\centering
	\def\@captype{figure}
	\makeatother
	\includegraphics[scale=1]{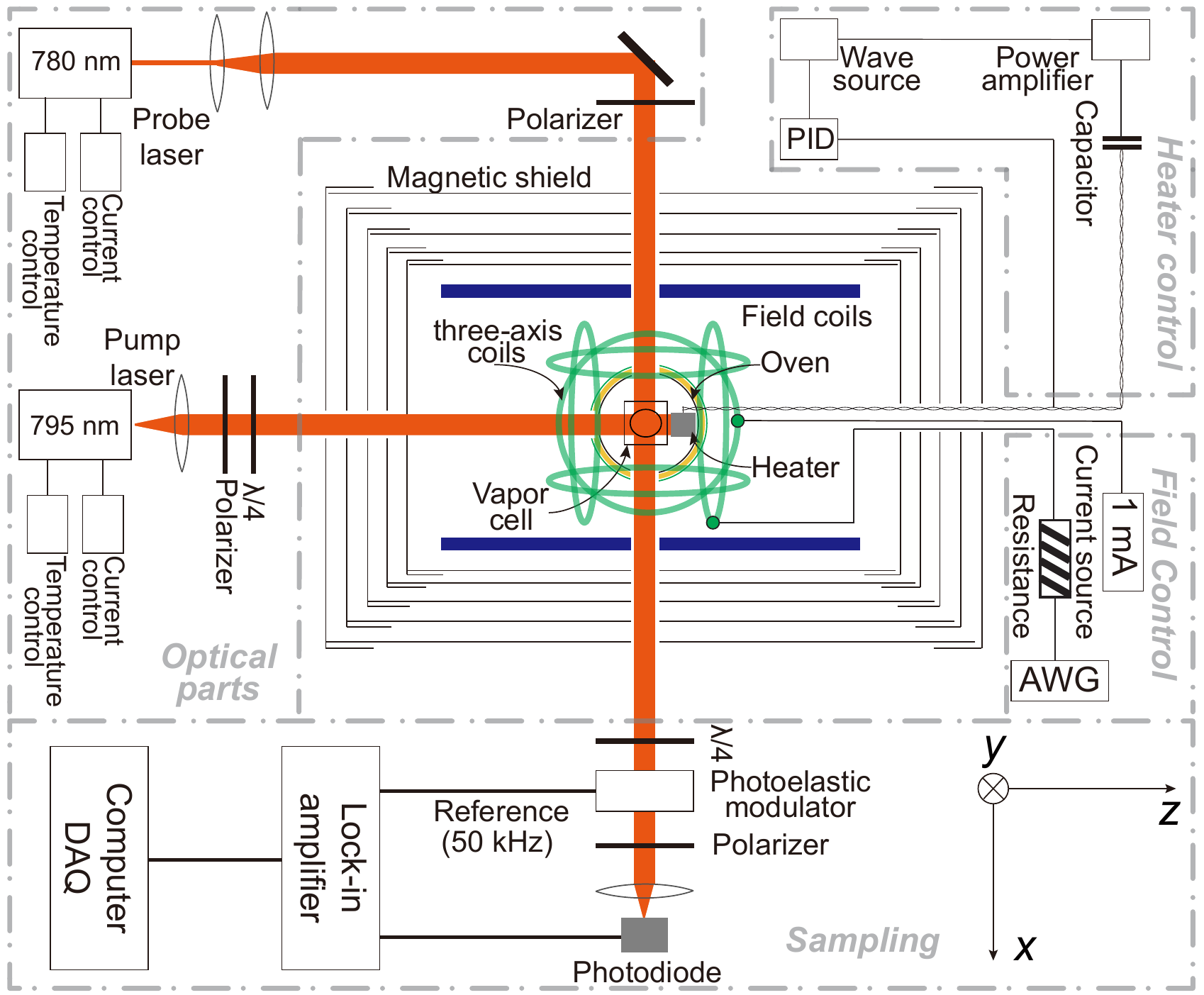}
	\caption{\textbf{Experimental setup for axion-like dark matter searches}. The details are present in the text. DAQ, data Acquisition; PID, proportional–integral–derivative controller; AWG, arbitrary wavefunction generator; $\lambda/4$, quarter-wave plate.}
	\label{figure1}
\end{figure*}

\section{Experimental apparatus}
This section presents the details of our experimental setup for axion-like dark matter searches.
Figure~\ref{figure1} shows the schematic of the experimental apparatus.
0.5~cm$^3$ volume cubic vapor cell made from pyrex glass contains 5~torr isotopically enriched $^{129}$Xe, 250~torr N$_2$ as buffer gas, and a droplet (several miligrams) of isotopically enriched $^{87}$Rb metal. The vapor cell is heated to $140$~$^\circ$C with twisted wire using AC current.
The cell is magnetically shielded with a five-layer cylindrical $\mu$-metal shield (shielding factor of 10$^6$). Two sets of three pairs orthogonal coils are placed around the vapor cell and can provide bias and oscillating magnetic fields in an arbitrary direction. A bias field $B_{0}^{z}$ (above 200~$\rm{nT}$) is applied along $z$ with current driven with a precision current source (Krohn-Hite Model 523) and an oscillating field $B_{\rm{ac}}^{y}$ is applied along $y$ with an arbitrary waveform generator (Keysight Model 33210A). The $^{87}$Rb atoms in the vapor cell are pumped with a circularly polarized laser light along ${z}$. The $^{129}$Xe atoms are polarized by spin-exchange collisions with optically pumped $^{87}$Rb atoms. The pump laser light is tuned to the D1 transition at 795~nm and its optical power is $\sim$26~$\textrm{mW}$. The magnetic field is measured via optical rotation of the linearly polarized probe laser beam propagating along the $x$ direction.
The probe beam is blue-detuned by 110~GHz from the D2 transition at 780~nm and its power is $\sim$0.8~$\textrm{mW}$.
The optical rotation of the probe beam after passing the vapor cell is
(see, for example, refs.~\cite{wu1986optical, opechowski1953magneto, jiang2020interference})
\begin{equation}
\theta=\frac{1}{4} lr_e cfn P_x^e D(v),
\label{FR}
\end{equation}
where $P_x^e$ is the electron spin polarization of $^{87}$Rb atoms along the $x$ axis;
its explicit form (which depends on the unknown magnetic fields) is described below,
$l\approx 8$~mm is the optical path length,
$r_e=2.8 \times 10^{-13}$~cm is the classical radius of the electron,
$c$ is the speed of light,
$f$ is the oscillator strength (about 1/3 for D1 light and 2/3 for D2 light),
$D(v)=(v-v_0)/[(v-v_0)^2+(\Delta v/2)^2]$,
$v$ is the frequency of the probe laser light,
and $\Delta v$ is the full-width at half-maximum (FWHM) of the optical D2 transition of frequency $v _0$.
To suppress the influence of low-frequency noise of the probe beam, the polarization of the probe beam is modulated with a photoelastic modulator (PEM) at 50~$\rm{kHz}$.
The signal is demodulated with a lock-in amplifier (SRS Model 830) and is acquired with a 24-bit acquisition card (NI 9239).

As discussed below, when an oscillating external magnetic field is resonant with $^{129}$Xe spins, the induced $^{129}$Xe transverse magnetization becomes considerable.
The transverse nuclear magnetization can generate an effective magnetic field on $^{87}$Rb atoms and
such magnetic field can be detected with a $^{87}$Rb magnetometer (the details are present in Sec.~\ref{sec2}).
We show below that the amplitude of the effective field could be significantly larger than that of the external oscillating field.
Thus, $^{129}$Xe spins act as a spin-based amplifier, see Sec.~\ref{sec2},
to boost the signal from the external oscillating magnetic field.
Using the $^{129}$Xe spin-based amplifier, the magnetic sensitivity of the $^{87}$Rb magnetometer can be enhanced by two orders of magnitude and can reach 18.5~fT/$\sqrt{\rm{Hz}}$.
This is in contrast to other state-of-the-art $^{87}$Rb magnetometers based on $^{87}$Rb-$^{129}$Xe vapor cells,
where their magnetic sensitivities are typically on the order of pT/$\sqrt{\rm{Hz}}$.
Benefiting from the high sensitivity of spin-based amplifier,
we use this device to search for axion-like dark matter, which can be seen as a coherently oscillating magnetic field (see Sec.~\ref{sec5B}).

\section{Analysis of spin-based amplifier}
\label{sec2}
This section presents a detailed analysis of our spin-based amplifier.
The deveice makes use of $^{129}$Xe noble gas, which spatially overlaps with $^{87}$Rb gas in the same vapor cell. The spin dynamics of the overlapping $^{129}$Xe and $^{87}$Rb can be described by the coupled Bloch equations~\cite{kornack2005nuclear},
\begin{eqnarray}
\label{H1}
\frac{\partial \textbf{P}^{e}}{\partial t}&=&\frac{\gamma_{e}}{Q} (\textbf{B}^{0}+\lambda {M}^{n}\textbf{P}^{n})\times\textbf{P}^{e} + \frac{P^{e}_{0} \bm{z} -\textbf{P}^{e} }{\{T_{2e},T_{2e},T_{1e}\} Q},\\
\label{H2}
\frac{\partial \textbf{P}^{n}}{\partial t}&=&\gamma_{n}(\textbf{B}^{0}+\lambda {M}^{e}\textbf{P}^{e})\times\textbf{P}^{n} +  \frac{P^{n}_{0}\bm{z}-\textbf{P}^{n} }{\{T_{2n},T_{2n},T_{1n}\}},
\end{eqnarray}
where $\gamma_{e}=2\pi\times 28~\textrm{Hz/nT}$ and $\gamma_{n}=2\pi\times 0.0118~\textrm{Hz/nT}$ are, respectively, the gyromagnetic ratio of a bare electron and $^{129}$Xe nuclear spin, $\textbf{P}^{e}$ and $\textbf{P}^{n}$ are, respectively, the polarization of $^{87}$Rb electrons and $^{129}$Xe atoms,  $P_{0}^{e}$ and $P_{0}^{n}$ are the equilibrium polarization of $^{87}$Rb and $^{129}$Xe, $M^{e}$ and $M^{n}$ are, respectively, the magnetization of $^{87}$Rb atoms and $^{129}$Xe atoms with unity polarization, $T_{1n},T_{2n} \approx 20$~s are the longitudinal and transverse relaxation times of $^{129}$Xe spins, $\textbf{B}^{0}$ is the applied magnetic field, and $\lambda M \textbf{P}$ is the effective field induced by $\textbf{P}$ polarization. $T_{1e},T_{2e}$ are the longitudinal and transverse relaxation times of $^{87}$Rb spins. The relaxation times of $^{87}$Rb spins satisfy $1/T_{2e}=1/T_{1e}+1/T_{2}^{\rm{SE}}$, where $1/T_{2}^{\rm{SE}}$ is the relaxation caused by spin-exchange collisions. In our experiment, $1/T_{1e}$ is the dominant relaxation term. Accordingly, $T_{1e} \approx T_{2e}$ can be regarded as the common relaxation time without distinction here. The factor $Q$ is the slowing-down factor of $^{87}$Rb atoms, which depends on the $^{87}$Rb polarization. Because $P_{z}^{e}$ is much larger than transverse $P_{x}^{e}$ and $P_{y}^{e}$, $Q$ primarily depends on $P_{z}^{e}$. Moreover, as $P_{z}^{e}$ is approximated as a constant, $Q$ can be approximated as a constant.

For spatially overlapping $^{87}$Rb and $^{129}$Xe spins, their Fermi-contact interaction introduces an effective field $\lambda M^{n} \textbf{P}^{n}$ on $^{87}$Rb spins generated by $^{129}$Xe nuclear magnetization~\cite{walker1997spin}. Such an effective field can be described by 
\begin{equation}
	\textbf{B}_{\rm{eff}}=\lambda M^{n} \textbf{P}^{n}=\frac{8\pi}{3} \kappa_0 M^{n} \textbf{P}^{n},
	\label{beff}
\end{equation}
 where the Fermi-contact enhancement factor $\kappa_{0}$ is about 540 and $\mu_{0}$ is the permeability of vacuum.
Similarly, $^{87}$Rb magnetization also generates an effective field $\lambda M^{e} \textbf{P}^{e}$ on $^{129}$Xe spins. In our experiment, $\lambda M^{e} \textbf{P}^{e}$ is on the order of nT and $\lambda M^{n} \textbf{P}^{n}$ is at least one order greater than $\lambda M^{e} \textbf{P}^{e}$. In Sec.~\ref{sec2C}, we present some numerical calculations of $\textbf{B}_{\rm{eff}}$.

We now consider how to solve the coupled Bloch equations (Eqs.~\ref{H1} and \ref{H2}). At first, we simplify them. Unlike the near-zero field case where noble gas spins and electron spins are strongly coupled together and operate as a comagnetometer~\cite{kornack2005nuclear}, we consider a different situation where $^{87}$Rb and $^{129}$Xe spins are under a large bias field $B_{0}^{z}$ along $z$ (above 200~nT).
In this situation, $^{87}$Rb and $^{129}$Xe spins are weakly coupled together, greatly simplifying Eqs.~\ref{H1} and \ref{H2}. The details are explained as follows. (1) $P^{n}_{z}$ can be approximated as a constant, which is much larger than $P^{n}_{x}$ and $P^{n}_{y}$. Based on Eq.~\ref{H2}, the $^{129}$Xe spins evolve under the static field of the effective field $\lambda M^{e} {P}_z^{e}$ along $z$ generated by $^{87}$Rb spins and the bias field $B^{z}_{0}$. Because the applied $B_{0}^{z}$ is much larger than the effective field $\lambda M^{e} {P}_z^{e}$, we can first neglect the term $\lambda M^{e} {P}_z^{e}$ in Eq.~\ref{H2} for simplicity (but the $^{87}$Rb effective field along $z$ should be considered when the Larmor frequency of $^{129}$Xe is calibrated, see Sec.~\ref{sec5A}). As a result, Eq.~\ref{H2} can be independently solved at first and $\textbf{P}^n$ can be evaluated. (2) Then, we can take the solution of $\textbf{P}^n$ into Eq.~\ref{H1} and solve for $\textbf{P}^e$.

We write the total magnetic field experienced by the $^{87}$Rb spins as $\textbf{B}=\textbf{B}^{0}+\lambda {M}^{n}\textbf{P}^{n}$ and obtain the simplified Bloch equations,
\begin{eqnarray}
\label{H3}
\frac{\partial \textbf{P}^{e}}{\partial t}&=&\frac{\gamma_{e}}{Q} \textbf{B} \times\textbf{P}^{e} + \frac{P^{e}_{0} \bm{z} -\textbf{P}^{e} }{T_{e} Q},\\
\label{H4}
\frac{\partial \textbf{P}^{n}}{\partial t}&=&\gamma_{n}\textbf{B}^{0}\times\textbf{P}^{n} +  \frac{P^{n}_{0}\bm{z}-\textbf{P}^{n} }{\{T_{2n},T_{2n},T_{1n}\}}.
\end{eqnarray}

We first solve the evolution of $^{129}$Xe spins under a bias field $B^{0}_{z}$ along $z$ and an oscillating magnetic field $B_{\textrm{ac}}^{y}\cos(2\pi\nu t)$. The Larmor frequency of $^{129}$Xe spins can be written as $\nu_{0}=\gamma_{n}B_{0}^{z}/(2\pi)$. With rotating-wave approximation, we can rewrite the Bloch equation of $^{129}$Xe spins in the rotating frame,

\begin{equation}
\begin{aligned}
\dfrac{\partial}{\partial t}\tilde{\textbf{P}}^{n}=\gamma_{n} \tilde{\textbf{B}}^{0} \times \tilde{\textbf{P}}^{n}-\dfrac{\tilde{P}_{x}^{n}\bm{x}+\tilde{P}_{y}^{n}\bm{y}}{T_{2n}}-\dfrac{(P_{z}^{n}-P_{0}^{n})\bm{z}}{T_{1n}},
\label{H6}
\end{aligned}
\end{equation}
where $\tilde{\textbf{B}}^0=\dfrac{2\pi(\nu_{0}-\nu)}{\gamma_{n}}\bm{z}+\dfrac{B_{\textrm{ac}}^{y}}{2}\bm{y}$ is the equivalent magnetic field in the rotating frame. We derive the three components of $\tilde{\textbf{P}}^{n}$ by solving Eq.~\ref{H6} in rotating frame and transforming back to the laboratory frame. Finally, the transverse polarization components are
\begin{equation}
\begin{aligned}
P^{n}_{x}=\dfrac{1}{2}P^{n}_{0} \gamma_{n} B_{\textrm{ac}}^{y}\dfrac{ T_{2n}\cos(2\pi\nu t)+2\pi(\nu-\nu_{0})T_{2n}^{2}\sin(2\pi\nu t)}{1+({\gamma_{n} B_{\textrm{ac}}^{y}}/2)^{2}T_{1n}T_{2n}+[2\pi(\nu-\nu_{0})]^{2}T_{2n}^{2}}+Ce^{-t/T_{2n}}\cos(2\pi\nu_{0} t),\\
P^{n}_{y}=\dfrac{1}{2}P^{n}_{0} \gamma_{n} B_{\textrm{ac}}^{y}\dfrac{T_{2n}\sin(2\pi\nu t)-2\pi(\nu-\nu_{0})T_{2n}^{2}\cos(2\pi\nu t)}{1+({\gamma_{n} B_{\textrm{ac}}^{y}}/2)^{2}T_{1n}T_{2n}+[2\pi(\nu-\nu_{0})]^{2}T_{2n}^{2}}+Ce^{-t/T_{2n} } \sin(2\pi\nu_{0} t),
\end{aligned}
\label{H9}
\end{equation}
where $C$ is a constant determined by initial conditions.
When the oscillating field $B^{y}_{\rm{ac}}$ is suddenly applied to the vapor cell, the transient response of $^{129}$Xe spins can be created, as described in Eq.~\ref{H9}.
According to $\textbf{B}_{\rm{eff}}=\lambda M^{n} \textbf{P}^{n}$, we can derive the effective field experienced by $^{87}$Rb atoms (or $^{87}$Rb magnetometer) as
\begin{equation}
\begin{aligned}
&\textbf{B}_{\rm{eff}}=\textbf{B}_{\rm{eff}}^{\rm{CW}}+\textbf{B}_{\rm{eff}}^{\rm{Tran}},\\
&\textbf{B}_{\rm{\rm{eff}}}^{\rm{CW}}=\overbrace{\dfrac{1}{2}\lambda M^{n} P^{n}_{0} \gamma_{n} B_{\textrm{ac}}^{y}\dfrac{T_{2n}\cos(2\pi\nu t)+2\pi(\nu-\nu_{0})T_{2n}^{2}\sin(2\pi\nu t)}{1+({\gamma_{n} B_{\textrm{ac}}^{y}}/2)^{2}T_{1n}T_{2n}+[2\pi(\nu-\nu_{0})]^{2}T_{2n}^{2}}\bm{x}}^\textrm{ effective field generated by $^{129}$Xe $x$ magnetization}\\
&~~~~~~~+\overbrace{\dfrac{1}{2}\lambda M^{n} P^{n}_{0} \gamma_{n} B_{\textrm{ac}}^{y}\dfrac{T_{2n}\sin(2\pi\nu t)-2\pi(\nu-\nu_{0})T_{2n}^{2}\cos(2\pi\nu t)}{1+({\gamma_{n} B_{\textrm{ac}}^{y}}/2)^{2}T_{1n}T_{2n}+[2\pi(\nu-\nu_{0})]^{2}T_{2n}^{2}}\bm{y}}^\textrm{effective field generated by $^{129}$Xe $y$ magnetization},\\
&\textbf{B}_{\rm{eff}}^{\rm{Tran}}=\overbrace{\lambda M^n C e^{-t/T_{2n}}\cos(2\pi\nu_{0} t)\bm{x}+ \lambda M^n C e^{-t/T_{2n} } \sin(2\pi\nu_{0} t)\bm{y}}^\textrm{transient response}.
\label{H7}
\end{aligned}
\end{equation}
To measure the steady-state response signal (i.e., $\textbf{B}_{\rm{\rm{eff}}}^{\rm{CW}}$) from $\textbf{B}_{\rm{eff}}$, we set 50~s waiting time to ensure that $\textbf{B}_{\rm{eff}}^{\textrm{Tran}}$ decays to zero and thus the transient term can be neglected before data sampling. Hence, in the following, $\textbf{B}_{\rm{eff}}^{\rm{Tran}}$ is neglected and $\textbf{B}_{\rm{eff}} \approx \textbf{B}_{\rm{eff}}^{\rm{CW}}$. The total oscillating magnetic field $\textbf{B}_{\rm{tot}}$ on $^{87}$Rb detected can be written as
\begin{equation}
\textbf{B}_{\rm{tot}}=\textbf{B}_{\rm{eff}}^{\textrm{CW}} +B^{y}_{\rm{ac}}\cos(2\pi\nu t) \bm{y}.
\label{H8}
\end{equation}

\begin{figure*}[t]  
	\makeatletter
\centering
	\def\@captype{figure}
	\makeatother
	\includegraphics[scale=1.3]{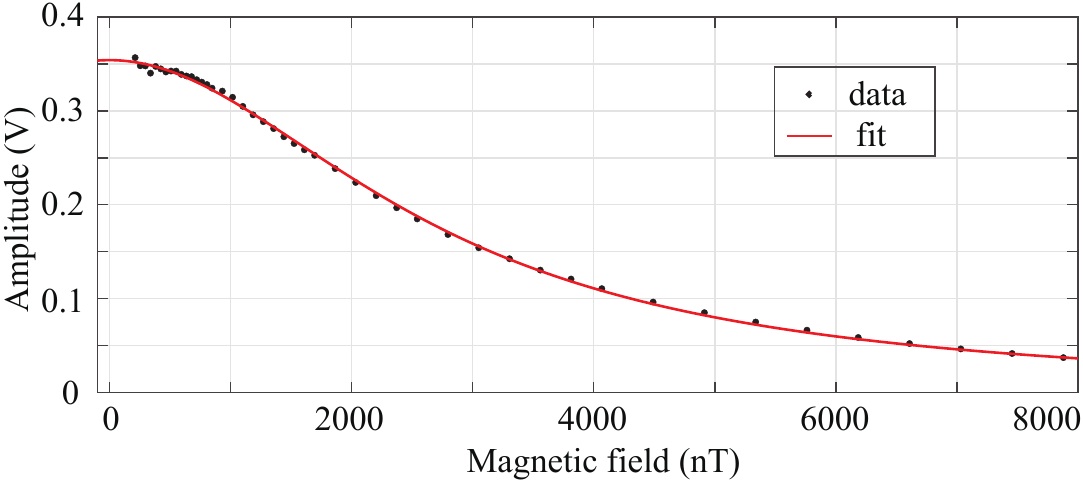}
	\caption{Response to an oscillating field ($B_y$) of the $^{87}$Rb magnetometer as a function of the bias magnetic field along with a fit to $\dfrac{-B_{y}{\Delta}B}{(B_z^0)^{2}+(\Delta B)^{2}}$, where the parameter is $\Delta B \approx 2700 \pm 13$~nT. In experiment, a 7.11~nT field with 320~Hz frequency is applied along ${y}$. The bias field $B^0_z$ is scanned from 0 to 8000~$\rm{nT}$, and the amplitude of the oscillating signal is recorded.}
	\label{response}
\end{figure*}

We now solve for the evolution of $^{87}$Rb spins (Eq.~\ref{H3}). In our experiment, the ${x}$ component of $^{87}$Rb polarization $P^{e}_{x}$ is detected with a probe beam (see Eq.~\ref{FR}). Thus, we only need obtain the explicit expression of $P^{e}_{x}$. Under the quasi-static field condition, we can obtain the steady-state solution
\begin{equation}
\begin{aligned}
P_{x} ^{e}\propto \dfrac{B_{x}B_{z}-B_{y}{\Delta}B}{|\textbf{B}|^{2}+(\Delta B)^{2}} \approx \dfrac{B_{x}B_{z}-B_{y}{\Delta}B}{(B_z^0)^{2}+(\Delta B)^{2}},
\label{H5}
\end{aligned}
\end{equation}
where ${\bigtriangleup}B=1/(\gamma_{e}T_{e})$. To calibrate the parameter ${\bigtriangleup}B$, we apply a 7.11~nT field with 320~Hz frequency along ${y}$ and scan the bias field from 0 to 8000~$\rm{nT}$. Figure~\ref{response} provides the experimental data, which is fitted with Eq.~\ref{H5} and yields $\Delta B \approx 2700 \pm 13$~nT. In our experiment, a static bias field $B^{z}_{0}$ above 200~nT is applied along the ${z}$ axis. In this situation, the $^{87}$Rb magnetometer becomes simultaneously sensitive to the magnetic fields along ${x}$ and ${y}$. This is in contrast to near-zero-field magnetometers, where $^{87}$Rb magnetometer is only sensitive to the magnetic field along $y$. Therefore, $^{87}$Rb magnetometer can measure the effective magnetic fields along ${x}$ and ${y}$ produced by $^{129}$Xe nuclear $x, y$-magnetization.


\subsection{Linear and nonlinear response}
\label{nonlinear}

The effective field $\textbf{B}_{\textrm{eff}}$ induced by $^{129}$Xe magnetization has a close relation with the measured $B_{\textrm{ac}}^y$ (for example, axion-like dark matter field). Thus, it is important to analyse the response of detectable $\textbf{B}_{\textrm{eff}}$ to unknown $B_{\textrm{ac}}^y$.
We consider three situations.

\begin{itemize}
\item[(1)] When the applied oscillating field $B^{y}_{\textrm{ac}}$ is weak and satisfies $({\gamma_{n} B_{\textrm{ac}}^{y}}/2)^{2}T_{1n}T_{2n} \ll 1$, based on Eq.~\ref{H7}, the effective field is proportional to the strength of $B^{y}_{\rm{ac}}$.
In this situation, the $^{129}$Xe spin-based amplifier has linear response to the measured $B_{\textrm{ac}}^y$.
According to our experimental parameters, the $B_{\textrm{ac}}^y$ should satisfy
\begin{equation}
	B^{y}_{\textrm{ac}} \ll \frac{2}{\gamma_n \sqrt{T_{1n} T_{2n}}} \approx 8.5~\textrm{nT}.
	\label{linearregime}
\end{equation}
\end{itemize}

\begin{itemize}
\item[(2)] When the applied oscillating field $B^{y}_{\rm{ac}}$ is comparable to that of Eq.~\ref{linearregime},
the $({\gamma_{n} B_{\textrm{ac}}^{y}}/2)^{2}T_{1n}T_{2n}$ term becomes significant.
In this situation, the response is nonlinear.
\end{itemize}

\begin{itemize}
\item[(3)] When the applied oscillating field $B^{y}_{\textrm{ac}}$ is much larger, the term $({\gamma_{n} B_{\textrm{ac}}^{y}}/2)^{2}T_{1n}T_{2n}$ cannot be neglected and causes nuclear magnetization-saturation effects. The effective field $\textbf{B}_{\rm{eff}}$ is negligible. In this situation, the linear response signal of the $^{87}$Rb magnetometer is dominant.
\end{itemize}

\begin{figure*}[h]  
	\makeatletter
\centering
	\def\@captype{figure}
	\makeatother
	\includegraphics[scale=1.5]{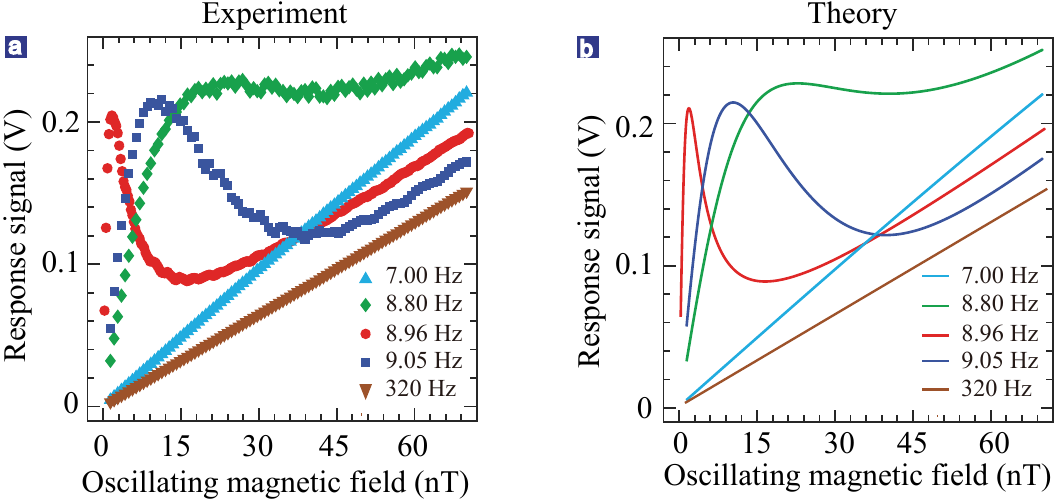}
	\caption{\textbf{Response of the spin-based amplifier as a function of the oscillating field amplitude}. (a) Experimental response as a function of oscillating magnetic field amplitude. Each experiment (with the oscillation frequency of 7.00, 8.80, 8.96, 9.05, 320~Hz, respectively) is performed in the same bias field $B_z^0 \approx 759$~nT (corresponding to that $^{129}$Xe Larmor frequency is 8.96~Hz). (b) Theoretical profiles corresponding to the experiment shown in (a). The experimental data agrees well with the theoretcal profiles.}
	\label{field}
\end{figure*}

We verify the above discussions (1)-(3) with experiments.
Figure~\ref{field}\textbf{a} provides experimental linear and nonlinear response signals,
scanned as a function of the oscillating field amplitude.
In contrast to the far off-resonant case (320 Hz),
the signals for resonant ($8.96$~Hz) and near-resonant cases ($8.80$~Hz, $9.05$~Hz) first linearly increase, then decrease, and lastly linearly increase again,
originating from nuclear magnetization saturation.
For example, the resonant signal (red circles) has a large response slope when the oscillating field amplitude is weak.
Due to the large slope, the signal from an external oscillating field can be enhanced.
With higher oscillating field amplitude,
the response signal decreases.
When the oscillating field amplitude is increased further, the response signal from the $^{87}$Rb magnetometer becomes linear again.
In the large oscillating field amplitude, the slopes for resonant ($8.96$~Hz) and near-resonant ($8.80$~Hz, $9.05$~Hz) cases are the same as that of the far off-resonant case (320 Hz).
This is because the response signals are from the rubidium magnetometer.
The solid curves in Fig.~\ref{field}\textbf{b} represent our theoretical calculations based on Eq.~\ref{H7}, which agree well with the experiment.



A particularly sensitive window for measuring external oscillating fields corresponds to the amplitude of oscillating fields below several nT (see Eq.~\ref{linearregime}).
In practice, we are interested in a sensitive measurement of small magnetic fields (for example, axion-like dark matter field),
whose amplitudes are within the sensitive window demonstrated in this work.
To maintain the operation in the sensitive window,
we should suppress ambient electromagnetic interference, for example, by using a five-layer magnetic shield and an ultralow-noise current source.

\begin{figure*}[h]  
	\makeatletter
\centering
	\def\@captype{figure}
	\makeatother
	\includegraphics[scale=1.2]{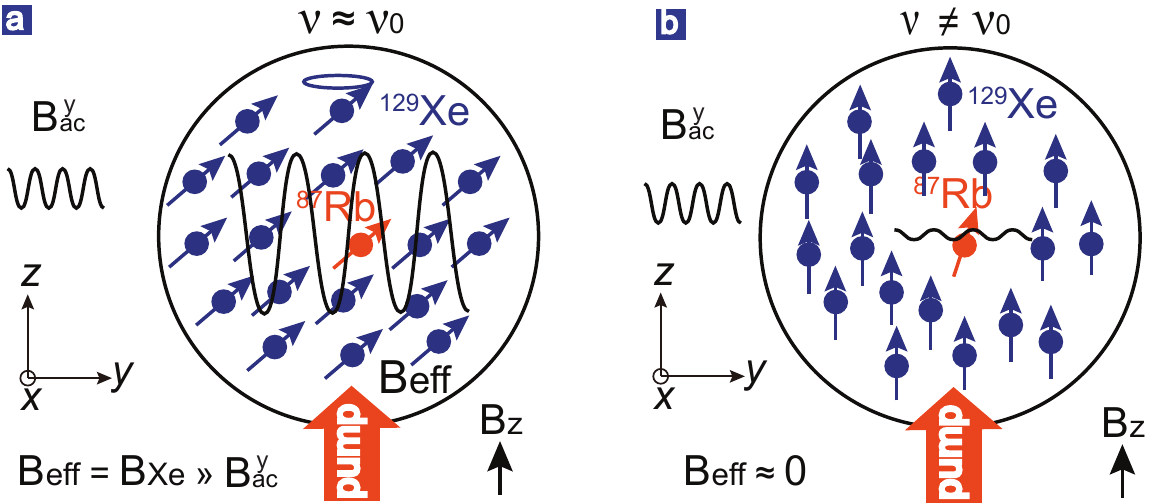}
	\caption{\textbf{Amplification performance of the spin-based amplifier}. (a) On- and near-resonance cases. When the frequency of the applied oscillating field $\nu$ is close to the $^{129}$Xe Larmor frequency $\nu_{0}$, the effective field $\textbf{B}_{\rm{eff}}$ is considerable. Such an effective field can be measured with $^{87}$Rb magnetometer. (b) Far-off-resonance cases. When the frequency of applied oscillating field $\nu$ is away from resonance frequency $\nu_{0}$, the effective field $\textbf{B}_{\rm{eff}}$ generated from $^{129}$Xe spins is negligible.}
	\label{principle}
\end{figure*}

Apart from the oscillating field amplitude,
the effective magnetic field $\textbf{B}_{\textrm{eff}}$ also depends on the oscillating field frequency.
We discuss the frequency dependence, as shown in Fig.~\ref{principle}:
(1) On-resonance case: when $\nu \approx \nu_{0}$, the effective field $\textbf{B}_{\rm{eff}}$ reaches a maximum and could be much larger than the oscillating field amplitude $B^{y}_{\rm{ac}}$.
We present how to calculate and experimentally measure the amplification factor,
as discussed in Sec.~\ref{sec2A};
(2) Near-resonance case: the effective field increases when the frequency of the oscillating field frequency $\nu$ becomes close to the Larmor frequency $\nu_{0}$.
Thus, there is a frequency bandwidth for the spin-based amplifier, as discussed in Sec.~\ref{sec2B}.
(3) Far-off-resonance case: when $\nu \gg \nu_{0}$ or $\nu \ll \nu_{0}$, the term $[2\pi(\nu-\nu_{0})]^{2}T_{2n}^{2}$ is dominant in Eq.~\ref{H7}. In this situation, the effective field $\textbf{B}_{\rm{eff}}$ generated by $^{129}$Xe spins is negligible and the applied oscillating field $B^{y}_{\rm{ac}}\cos(2\pi\nu t) \bm{y}$ is dominant.

\subsection{Amplification factor}
\label{sec2A}

The resonant oscillating field can induce an oscillating $^{129}$Xe nuclear magnetization, which can generate a considerable effective magnetic field $\textbf{B}_{\rm{eff}}$ on $^{87}$Rb spins.
As discussed below, the strength of $\textbf{B}_{\rm{eff}}$ could be much larger than the applied oscillating field $B^{y}_{\rm{ac}}$.
Accordingly, we define an amplification factor:
\begin{equation}
\eta=|\textbf{B}_{\rm{eff}}/B^{y}_{\rm{ac}}|.
\end{equation}

\begin{figure*}[h]  
	\makeatletter
\centering
	\def\@captype{figure}
	\makeatother
	\includegraphics[scale=1.5]{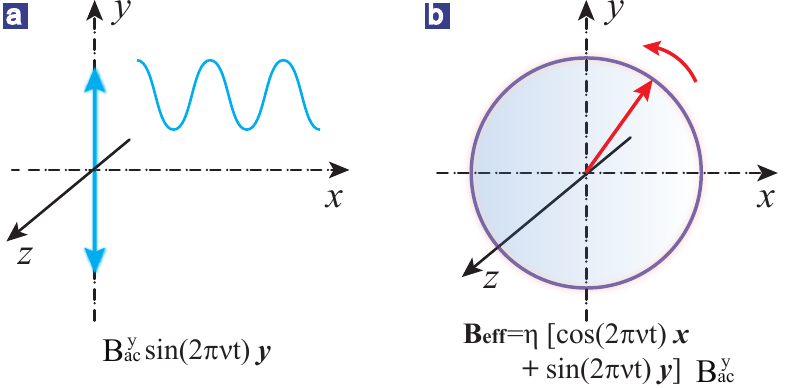}
	\caption{\textbf{Effective magnetic field $\textbf{B}_{\textrm{eff}}$}. (a) An oscillating field $B_{\textrm{ac}}^y \bm{y}$ is applied along the $y$ axis. (b) The magnetic field generated from $^{129}$Xe magnetization traces a circle in the $xy$ plane.}
	\label{eff_field}
\end{figure*}

We first derive the amplification factor $\eta$ on resonance. Here we assume that the oscillating field strength is small and thus the term $({\gamma_{n} B_{\textrm{ac}}^{y}}/2)^{2} T_{1n} T_{2n}$ can be neglected in Eq.~\ref{H7}. In this situation, the spin-based amplifier works in the sensitive linear-response regime and $\textbf{B}_{\rm{eff}}$ can be written as
\begin{equation}
	 \textbf{B}_{\textrm{eff}}(\nu=\nu_0)=\dfrac{1}{2}\lambda M^{n} P^{n}_{0} \gamma_{n} T_{2n}[\cos(2\pi\nu t)\bm{x} + \sin(2\pi\nu t)\bm{y}]B_{\textrm{ac}}^{y}.
	 \label{beff}
\end{equation}
Based on this result, the effective field $\textbf{B}_{\textrm{eff}}$ is a circularly polarized field (see Fig.~\ref{eff_field}) and its amplitude is equal to $\dfrac{1}{2}\lambda M^{n} P^{n}_{0} \gamma_{n} T_{2n} \cdot B_{\textrm{ac}}^y$.
As a result, the amplification factor is
\begin{equation}
	\eta=\dfrac{1}{2}\lambda M^{n} P^{n}_{0} \gamma_{n} T_{2n}.
	\label{eta}
\end{equation}
Based on Eq.~\ref{eta}, many ways can be used to increase the amplification factor, such as prolonging relaxation time $T_{2n}$ and improving equilibrium polarization $P^{n}_{0}$. The amplification factor can be as large as 10$^4$ in a $^3$He-K magnetometer (see Sec.~\ref{sec2C} for details). The sensitivity of $^3$He-K magnetometer can be improved by four orders of magnitude and can potentially reach a few aT/$\sqrt{\rm{Hz}}$.

Although the amplification factor can be theoretically calculated based on Eq.~\ref{eta},
this requires knowledge of experimental parameters $\{M^n, P_0^n, T_{2n}\}$.
Alternatively, we provide a simple experimental method to measure the amplification factor $\eta$.
The method consists of the following steps:

\begin{itemize}
\item[(1)] A resonant oscillating field is applied along $y$;
the output signal of $^{87}$Rb magnetometer is an oscillating voltage signal and we record its amplitude as $A (B_\textbf{eff})$.
By taking $\textbf{B}_{\textrm{eff}}$ (see Eq.~\ref{beff}) into $P_x^e$ (Eq.~\ref{H5}),
we achieve the output signal of the $^{87}$Rb magnetometer
\begin{equation}
\begin{aligned}
A (B_\textbf{eff}) \propto P^e_x(\textbf{B}_{\textrm{eff}})\propto \dfrac{1}{2} \lambda M ^{n}P_{0}^{n} \gamma_{n} T_{2n}\dfrac{1}{\sqrt{[{(B^{0}_{z})^{2}}+({\Delta B})^2]}}B_{\rm{ac}}^{y}.
\label{G1}
\end{aligned}
\end{equation}
In a practical measurement, we usually do not know the exact Larmor frequency without prior calibration. In this situation, an alternative way is to scan the oscillating field frequency over a small frequency range,
corresponding to the narrow bandwidth of the spin-based amplifier (see Sec.~\ref{sec2B}). The oscillating field frequency is automatically changed through an AWG (keysight 33210A) with a 0.04~Hz step applied. We fit the experimental data and find the maximum as the resonant signal amplitude.
\end{itemize}

\begin{itemize}
\item[(2)] A far-off-resonant oscillating field is applied along $y$; similarly, we record the amplitude $A (B_\textbf{ac}^y \textbf{y})$ of the $^{87}$Rb magnetometer output signal.
In this situation, we only need to consider $B^y_{\textrm{ac}}\bm{y}$ without considering $\textbf{B}_{\textrm{eff}}$.
The direct response of the $^{87}$Rb magnetometer to $B^y_{\textrm{ac}}\bm{y}$ can be expressed as
\begin{equation}
\begin{aligned}
A (B_\textbf{ac}^y \bm{y}) \propto P^e_x(B^y_{\textrm{ac}}\bm{y})\propto \dfrac{\Delta B}{[{(B^{0}_{z})^{2}}+({\Delta B})^2]}B_{\rm{ac}}^{y}.
\label{G2}
\end{aligned}
\end{equation}
Here, the frequency of the far-off-resonant field should be within the bandwidth of the $^{87}$Rb magnetometer, otherwise a frequency-response factor should be taken into account.
\end{itemize}

\begin{figure*}[b]  
	\makeatletter
\centering
	\def\@captype{figure}
	\makeatother
	\includegraphics[scale=1.5]{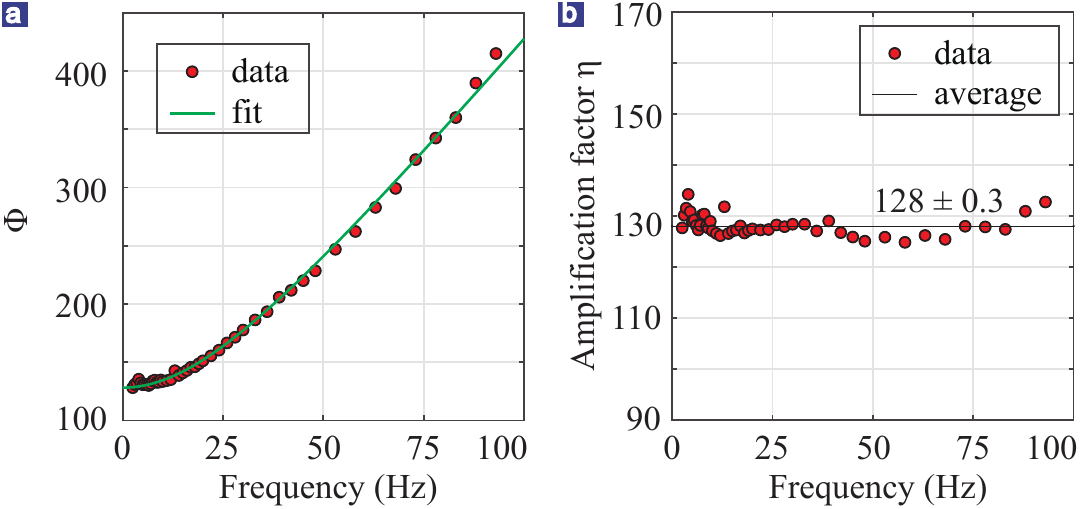}
	\caption{\textbf{Measurements of amplification factor}. (a) $\Phi$ as a function of resonant frequency (corresponding to different applied bias field). The solid line is a fit based on Eq.~\ref{G3} with $\eta \approx 128.1$. (b) The measured amplification factor $\eta$ at different resonant frequencies. The averaged $\eta$ is measured to be $\eta=128 \pm 0.3$.}
	\label{amplification}
\end{figure*}

\begin{itemize}
\item[(3)] By comparing the above two amplitudes, we have
\begin{equation}
\begin{aligned}
\Phi (\nu_0) &= \dfrac{A (B_\textbf{eff})}{A (B_\textbf{ac}^y \bm{y})}= \dfrac{1}{2}\lambda M ^{n}P_{0}^{n} \gamma_{n} T_{2n} \cdot \sqrt{1+\left( \dfrac{\nu_0}{\gamma_n \Delta B} \right)^{2}},
\label{G3}
\end{aligned}
\end{equation}
and further, the amplification factor can be evaluated by the following relation
\begin{equation}
	\begin{aligned}
	\eta  &= \Phi (\nu_0)/\sqrt{1+\left( \dfrac{\nu_0}{\gamma_n \Delta B} \right)^{2}}.
	\end{aligned}
	\label{eta1}
\end{equation}

Figure~\ref{amplification}\textbf{a} shows the experimental $\Phi (\nu_0)$ as a function of the resonant frequency, which agrees well with our theoretical model (shown with green line) described in Eq.~\ref{G3}.
The amplitude $A (B_\textbf{eff})$ scales as $1/B_z^0$,
while the amplitude of Eq.~\ref{G2} scales as $1/(B_z^0)^2$.
This is eventually leads to the dependence on Fig.~\ref{amplification}\textbf{a}.
The amplification factors calculated according to Eq.~\ref{eta1} are shown in Fig.~\ref{amplification}\textbf{b}.
The experimental result shows that the achieved amplification factors are nearly independent of $\nu_0$, which agrees with our theoretical model.
The average of such amplification factors is $128 \pm 0.3$.

\end{itemize}

In the axion-like dark matter searches (Secs.~\ref{sec4} and \ref{sec5}), we focus on the search via the axion-nucleon interactions known as `the axion gradient effect'~\cite{kimball2020overview, abel2017search, wu2019search, smorra2019direct, garcon2019constraints, graham2018spin, graham2013new} (see Sec.~\ref{sec4A}). In this model, axion-like dark matter acts as a time-oscillating magnetic field which couples to nuclear spins. Through spin-based amplifier, the effect of axion-like field can be enhanced and the measurement sensitivity can be greatly improved.

\subsection{Bandwidth}
\label{sec2B}
\begin{figure*}[t]  
	\makeatletter
\centering
	\def\@captype{figure}
	\makeatother
	\includegraphics[scale=1.3]{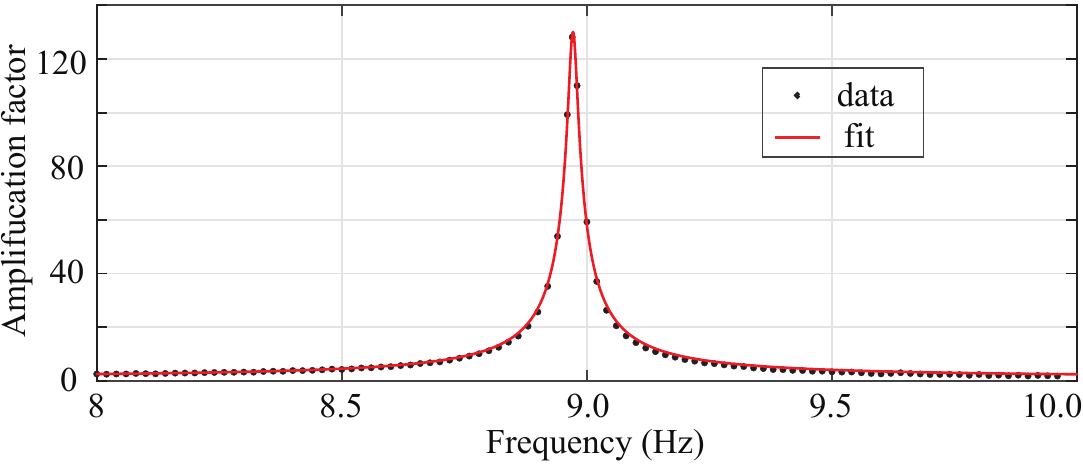}
	\caption{\textbf{Bandwidth measurement}. With a fit to $A/\sqrt{(\nu-\nu_{0})^2+(\Lambda /2)^2}$, where the bandwidth $ \sqrt{3}\Lambda =$ 0.052~Hz.}
	\label{frequency}
\end{figure*}

In Sec.~\ref{sec2A}, we only consider the amplification factor in the resonant case. We now analyse the amplification performance in the near-resonant case.
Based on the analysis of the frequency dependence above, the amplification factor reaches a maximum on resonance, and rapidly decreases when the frequency of applied oscillating magnetic field $\nu$ is far off-resonance. Therefore, there is a bandwidth for spin-based amplifier. Based on Eq.~\ref{H7}, we can rewrite the equation in the amplitude spectrum,

\begin{equation}
\begin{aligned}
|\textbf{B}_{\textrm{eff}}^{\textrm{CW}}| \propto \dfrac{\Lambda/2}{\sqrt{(\nu-\nu_{0})^2+(\Lambda/2)^2}}.
\label{G4}
\end{aligned}
\end{equation}
We assume that the oscillating field amplitude is weak and thus the term $({\gamma_{n} B_{\textrm{ac}}^{y}}/2)^{2}T_{1n}T_{2n}$ can be neglected in Eq.~\ref{H7}. As a demonstration, we set the bias field as $B_{0}^{z}=759$~nT and the oscillating field strength $B_{\textrm{ac}}^y=30$~pT. The frequency response is measured from 8 to 10 Hz, as shown in Fig.~\ref{frequency}. The experimental data are well described as a single-pole, band-pass filter (solid red curve). The full-width at half-maximum (FWHM) is $ \sqrt{3} \Lambda \approx 0.052$~Hz; the spin-based amplifier can enhance the signal in a correspondingly narrow frequency range.

\section{Derivation of the effective magnetic field induced by axion-like dark matter}
\label{sec4}
This section derives the effective magnetic field induced by axion-like dark matter.

\subsection{Axion dark-matter halo}
\label{sec4A}

Axion-like particles could be hypothetically produced in the early Universe by non-thermal mechanisms, such as ``vacuum misalignment"~\cite{marsh2016axion}. Then, they form a coherently oscillating classical field $a(t)=a_{0}\cos(2\pi\nu_{a} t)$~\cite{kimball2020overview}. The axion-like dark matter field $a(t)$ oscillates at its Compton frequency
\begin{equation}
\begin{aligned}
&\nu_{a}=\dfrac{m_{a}c^{2}}{h},
\label{J1}
\end{aligned}
\end{equation}
where $m_a$ is the mass of the axion-like particle, $c$ is the speed of light, and $h$ is the Planck constant. The coherence time determined by the random fluctuations of the field corresponds to the de Broglie wavelength $\lambda_{\rm{dB}}$,
\begin{equation}
\begin{aligned}
\tau_{a}\approx \dfrac{\lambda_{\rm{dB}}}{v}\approx \dfrac{h}{m_{a}v^{2}},
\label{J2}
\end{aligned}
\end{equation}
where $|\bm{v}| \sim 10^{-3}c$ is the galactic virial velocity of the axion dark-matter field. Assuming that axion field comprises the totality of the local dark matter energy density $\rho_{\rm{DM}}\approx 0.4~\rm GeV /cm^{3}$, the amplitude $a_{0}$ of the field obeys the equation,
\begin{equation}
\begin{aligned}
\rho_{\rm{DM}}\approx \dfrac{c^2}{2 \hbar^{2}} m_{a}^{2} a_{0}^{2}.
\label{J3}
\end{aligned}
\end{equation}
When axion-like dark matter field interacts with nuclear spins, the axion-like dark matter can be considered as a time-oscillating pseudo-magnetic field directed along the $\bm{v}$. Thus, the dark-matter couplings to nuclear spins can be seen as a Zeeman interaction between nuclear spins and this pseudo-magnetic field generated by axion-like particles. The resultant Hamiltonian is~\cite{kimball2020overview, abel2017search, wu2019search, smorra2019direct, garcon2019constraints, graham2018spin, graham2013new}
\begin{equation}
\begin{aligned}
\mathcal{H}_{\rm{spin}}
\approx \textrm{g}_{\rm{aNN}} \sqrt{2 \hbar^3 c \rho_{\rm{DM}}} \sin(2\pi \nu_{a}t) \bm{v} \cdot \textbf{I}_{\rm{N}},
\end{aligned}
\label{J7}
\end{equation}
where $\textrm{g}_{\rm{aNN}}$ is the strength of the axion-nucleon coupling and $\textbf{I}_{\rm{N}}$ is the spin-matrix vector of the nuclear spin. For a nucleus with a particular gyromagnetic ratio $\gamma_{n}=g_{n}\mu_{\rm{N}}/\hbar$, the $g_{n}$ is the nuclear Land$\acute{e}$ factor and $\mu_{\rm{N}}$ is the nuclear magneton. As a result, we can derive the effective magnetic field induced by axion-like dark matter,
\begin{equation}
\begin{aligned}
&\textbf{B}_{a} \approx 10^{-3} \dfrac{\textrm{g}_{\rm{aNN}}}{\hbar \gamma_{n}} \sqrt{2 \hbar^{3} c^3 \rho_{\rm{DM}}}\sin(2\pi \nu_{a}t) \times \dfrac{\bm{v}}{|\bm{v}|},\\
&\textbf{B}_{a} [\textrm{T}]\approx 10^{-7} \dfrac{\rm{g}_{\rm{aNN}}[\rm{GeV}^{-1}]}{\rm{g}_{n}}\sin(2\pi \nu_{a}t) \times \dfrac{\bm{v}}{|\bm{v}|}.
\end{aligned}
\label{J9}
\end{equation}

According to Eq.~\ref{J9}, we can build the relation between the detection of magnetic field and the coupling strength $\textrm{g}_{\rm{aNN}}$. Here, $\textrm{g}_{\rm{aNN}}$ is the important parameter that we should measure or constrain in experiments. In Sec.~\ref{sec5}, we present the details how to use our experimental data to constrain the coupling strength $\textrm{g}_{\rm{aNN}}$ over an axion mass range, from 10~feV to 1~peV or 2~Hz to 200~Hz.

\subsection{Coordinate transformation}
\label{sec4B}

\begin{figure*}[t]  
	\makeatletter
\centering
	\def\@captype{figure}
	\makeatother
	\includegraphics[scale=1.45]{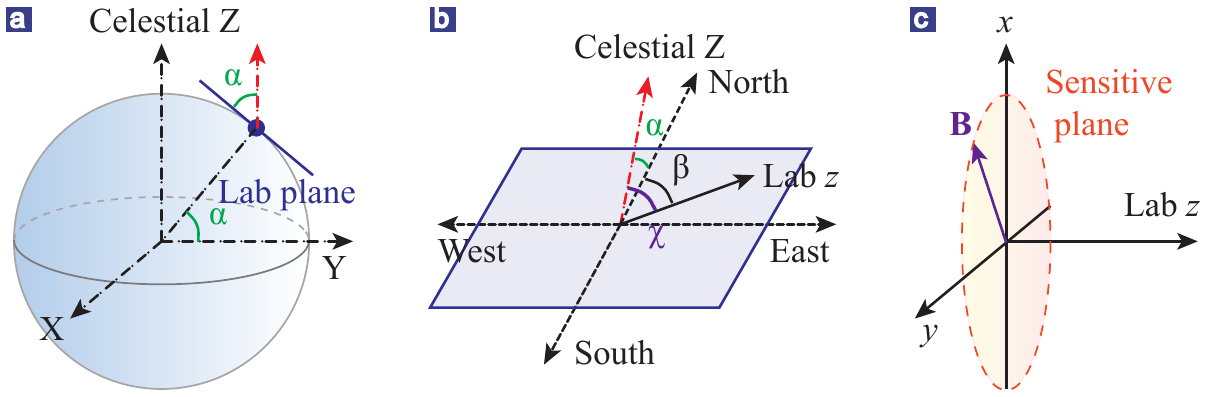}
	\caption{\textbf{Coordinates transformation}. (a) Relative relation between the celestial coordinates and the lab plane in celestial coordinates. (b) Relation between the celestial North and the lab coordinates. (c) The magnetic sensitive plane of the spin-based amplifier.}
	\label{coordinates}
\end{figure*}

Based on Eq.~\ref{J9}, the effective magnetic field is along the galactic virial velocity $\bm{v}$. For terrestrial experiments, we need to calculate the projection component of $\textbf{B}_{a}$ along the sensitive axis of our setup shown in Fig.~\ref{coordinates}$\textbf{b}$. To this end, we should calculate the projection of the virial velocity along sensitive axis. In lab coordinates, $^{129}$Xe spins are polarized along the ${z}$ axis and the transverse magnetization is detected. As discussed in Sec.~\ref{sec2}, the sensitive direction of the spin-amplifier-based magnetometer is perpendicular to the bias field (along ${z}$), see Fig.~\ref{coordinates}$\textbf{c}$. Thus, we need to calculate the projection of pseudo-magnetic field in the plane perpendicular to the ${z}$ axis in laboratory frame.

There are three coordinate systems, which are galactic coordinates, the celestial coordinates and laboratory coordinates. The virial velocity direction is defined as a direction of 270$^\circ$ longitude and 0$^\circ$ latitude in galactic coordinates. We use the coordinate convert tool from the official NASA website (NASA LAMBDA--Tools), to convert the direction of the axion velocity from galactic coordinates to celestial coordinates. The axion velocity in celestial coordinates can be written as $\bm{v}=\vert \bm{v} \vert (\cos \delta \cos \eta \bm{X}+ \cos \delta \sin \eta \bm{Y}+ \sin \delta \bm{Z})$,
where declination is $\delta=-48^\circ$ and right ascension is $\eta=138^\circ$~\cite{wu2019search, smorra2019direct}.

Next, we transform the axion velocity in celestial coordinates into laboratory coordinates. Based on ref.~\cite{kostelecky1999constraints}, we can use the following matrix to convert coordinates,

\begin{equation}
\begin{bmatrix}
x\\y\\z
\end{bmatrix}
=
\begin{bmatrix}
\cos \chi \cos \Omega t &\cos \chi \sin \Omega t & -\sin \chi
\\-\sin \Omega t& \cos \Omega t &0 \\ \sin \chi \cos \Omega t &\sin \chi \sin \Omega t & \cos \chi
\end{bmatrix}
\begin{bmatrix}
X\\Y\\Z
\end{bmatrix},
\label{K1}
\end{equation}
where $\begin{bmatrix} x&y&z \end{bmatrix}^{\rm{T}}$ and $\begin{bmatrix} X&Y&Z \end{bmatrix}^{\rm{T}}$ are lab and celestial coordinates, $\Omega=2\pi \times 1.16 \times 10^{-5}$~s$^{-1}$ is the sidereal frequency or the Earth's rotation frequency, $\chi$ is the angle between the ${z}$ axis in lab and in celestial coordinates. We can derive this angle through the equation $\cos \chi =\cos \alpha \cos \beta$, where $\alpha$ is the latitude of our lab and $\beta$ is the angle between the ${z}$ axis of lab and the Geographic North (the axis ${z}$ of celestial coordinates projected in the lab plane). Our apparatus for axion-like dark matter searches is located in University of Science and Technology of China (Hefei, China), where $\alpha=30^{\circ}$ and $\beta=90^{\circ}$. From the matrix Eq.~\ref{K1}, we can obtain the relation between coordinates transformation $z=\sin \chi \cos \Omega t X+ \sin \chi \sin \Omega t Y +\cos \chi Z $.
Finally, we derive the axion velocity $v_{z}$ in lab coordinates,

\begin{equation}
v_{z}=\bm{v} \cdot \bm{z}= \vert \bm{v} \vert [\cos \chi \sin \delta +\sin \chi \cos \delta \cos(\Omega t -\eta)].
\end{equation}
Because $\cos \chi $ is close to 0 based on the relation $\cos \chi =\cos \alpha \cos \beta$, we can ignore the $\cos \chi \sin \delta$ term. Based on this simplification,
we derive the transverse velocity:

\begin{equation}
\sqrt{v_{x}^2+v_{y}^{2}}=\sqrt{\bm{v}^{2}-v_{z}^{2}}=\vert \bm{v} \vert \sqrt{1-[\sin \chi \cos \delta \cos^{2}(\Omega t- \eta)]}=\vert \bm{v} \vert \sqrt{1-3/4\cos^{2}(\Omega t- \eta)}.
\end{equation}

Substituting the velocity term back into Eq.~\ref{J7}, we derive the equation utilized to search for axion-like dark matter in our experiment:
\begin{equation}
{B}_{a}[\rm{T}] \approx 10^{-7} \dfrac{\textrm{g}_{\rm{aNN}}[\rm {GeV}^{-1}]}{g_{n}} \sin(2\pi \nu_{a} t) \times \sqrt{1-3/4\cos^{2}(\Omega t- \eta)}.
\label{J8}
\end{equation}
According to Eq.~\ref{J8}, axion-like dark matter induced pseudo-magnetic field is modulated by the Earth's rotation. Because the sampling time for each search measurement is much shorter than the Earth's rotation period, $\sqrt{1-3/4\cos^{2}(\Omega t- \eta)}$ during the sampling time can be appromixated as the value at the initial time of each search experiment.

\section{Data analysis for the axion-like dark matter search}
\label{sec5}
This section presents the detailed experimental procedure and data analysis of axion-like dark matter searches. Our search experiment includes two parts,
i.e., calibration and axion search.

\subsection{Calibration experiments}
\label{sec5A}

\begin{figure*}[t]  
	\makeatletter
\centering
	\def\@captype{figure}
	\makeatother
	\includegraphics[scale=1.4]{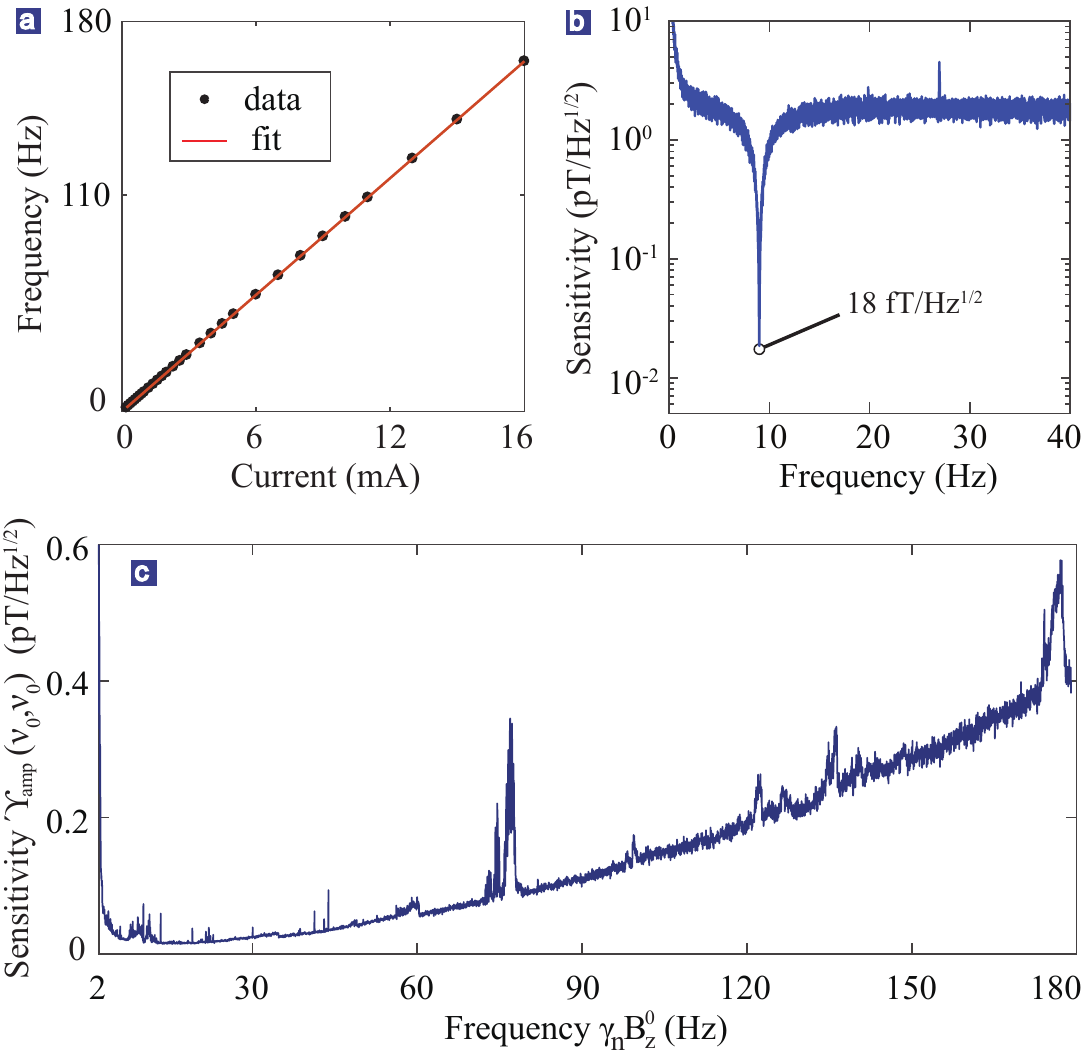}
	\caption{\textbf{Results of experimental calibrations.} (a) Calibration of relation between $^{129}$Xe Larmor frequency and applied coil current. (b) Calibration of $^{87}$Rb magnetometer magnetic sensitivity assisted the $^{129}$Xe spin-based amplifier. Note that $18~\textrm{fT} / \textrm{Hz}^{1/2}$ is achieved at $^{129}$Xe Larmor frequency ($\nu_0 \approx 8.96$~Hz), which is beyond the sensitivity ($\approx$2~pT/Hz$^{1/2}$) of the rubidium magnetometer itself. (c) Calibration of the enhanced magnetic sensitivity over different bias fields. There are some spurious noise peaks (e.g., around 80~Hz and 180~Hz) that originate from the probe-beam intensity noise, which can be removed by properly tuning and stabilizing the probe laser.}
	\label{calibaration}
\end{figure*}

As demonstrated in Sec.~\ref{sec2B}, the $^{87}$Rb magnetometer assisted with $^{129}$Xe spin-based amplifier can be highly sensitive to the oscillating magnetic field whose frequency is within the bandwidth of the spin-based amplifier. To use such a magnetometer in the searches of axion-like dark matter, it is the prerequisite to calibrate the frequency bandwidth of magnetometer and the magnetic sensitivity over the bandwidth.
To this end, we should perform three calibration experiments: 


\begin{itemize}
\item[(1)] Calibration of the relation between $^{129}$Xe Larmor frequency and applied coil current.
The bias magnetic field is generated with a commercial current source (Krohn-Hite Model 523). This current source drives a set of coils (along $z$). Due to the existence of $^{87}$Rb effective field (see Sec.~\ref{sec2}) and magnetic shield residual field, we need to carefully calibrate the relation between $^{129}$Xe Larmor frequency and the applied current.
At first, to measure the $^{129}$Xe Larmor frequency, we apply a $\pi /2$ magnetic-field pulse along the ${x}$ axis to rotate the $^{129}$Xe spins from ${z}$ axis to ${y}$ axis, and then the free-decay of $^{129}$Xe signals are measured for 100~s.
The free-decay signal can be fitted with a single-exponential-decay trigonometric function,
which gives corresponding Larmor frequency.
Figure~\ref{calibaration}$\textbf{a}$ shows experimental Larmor frequencies (black circles) as a function of the applied current.
The linear fit for the experimental data yields

\begin{equation}
\nu_0 (I)=\frac{\gamma_n}{2\pi} (761 \cdot I +0.66),
\end{equation}
where $I$ is the applied current (mA) and the conversion coefficient between magnetic field and applied current is about $\approx 761$~nT/mA.
The intercept of 0.66~nT represents the sum of $^{87}$Rb effective field and magnetic shield residual field.
Based on our calibration, we determine the dark-matter search frequency range when the applied current is set.

\end{itemize}

\begin{itemize}
\item[(2)] Calibration of $^{87}$Rb magnetometer magnetic sensitivity over different bias fields. Here the amplification effect is not taken into account and we only calibrate the $^{87}$Rb magnetometer sensitivity itself.
A certain bias field is first set and then we wait for 50~s before 100~s sampling.
The reason of 50-s waiting is to make sure that the set current becomes stable.
During the procedure, a 320~Hz calibration field with 7.11~nT amplitude is applied along $y$.
Based on the information of the calibration field, we determine the conversion coefficient between voltage (the $^{87}$Rb magnetometer output signal is voltage) and applied oscillating field.
Then, we calculate the power spectrum of 100~s data,
which can be converted to magnetic field sensitivity.
In particular, for each bias field $B^0_z$, we record the $^{87}$Rb magnetometer sensitivity at the corresponding $^{129}$Xe Larmor frequency $\nu_0=\gamma_n B^0_z$, which is the most sensitive frequency because of spin-based amplifier (see Sec.~\ref{sec2B}). Further, we scan bias fields to repeat the above step. In a small bias field, for example at $B^0_z=1125$~nT, the magnetic sensitivity can reach $\Upsilon_{\rm{Rb}} (\gamma_n B^0_z \approx 13.3) \approx 1.5~\rm{pT}/\sqrt{\rm{Hz}}$.
With increasing bias field $B^0_z$, the sensitivity gradually decreases due to the loss of $\frac{\partial  P_x^e}{\partial  B_y} \propto [(B^0_z)^2+(\Delta B)^2]^{-1}$ (see Eq.~\ref{H5}).
In a large bias field, for example at $B^0_z= 15000$~nT, the sensitivity is $\Upsilon_{\rm{Rb}} (\gamma_n B^0_z \approx 177.0) \approx 58~\rm{pT}/\sqrt{\rm{Hz}}$.
\end{itemize}

\begin{itemize}
\item[(3)] Calibration of amplification factor and enhanced magnetic sensitivity.
The detailed procedure to measure amplification factor is described in Sec.~\ref{sec2A}.
The amplification factor is measured to be $\eta \approx 128$.
Further, we can calculate the enhanced magnetic sensitivity at scanned Larmor frequency

\begin{equation}
\Upsilon_{\rm{amp}}(\nu_0,\nu_{0})=\frac{\Upsilon_{\rm{Rb}}(\nu_{0})} {\Phi(\nu_0)},
\label{P10}
\end{equation}
where $\Phi(\nu_0)$ represents the ratio between the resonant response signal amplitude and far off-resonant response signal amplitude, as seen in Eq.~\ref{eta1}:
\begin{equation*}
\Phi(\nu_0)=\eta \sqrt{1+\left(\frac{\nu_0}{\gamma_n \Delta B}\right)^2}.
\end{equation*}
Figure~\ref{amplification}\textbf{a} shows the experimental result for $\Phi(\nu_0)$.
Figure~\ref{calibaration}$\textbf{b}$ signals the significant improvement of magnetic sensitivity of the spin-based amplifier,
comparing with the $^{87}$Rb magnetometer itself.
Figure~\ref{calibaration}$\textbf{c}$ shows the $^{87}$Rb magnetometer magnetic sensitivity at $^{129}$Xe Larmor frequency over different bias fields.
However, the scanned Larmor frequency is discrete.
For other frequencies $\nu$ nearby each resonant frequency $\nu_0$,
as demonstrated in Sec.~\ref{sec2B}, the spin-based amplifier can operate within the bandwidth of 0.052~Hz centered at $^{129}$Xe Larmor frequency.
Therefore, it is necessary to derive the explicit form of the magnetic sensitivity of the spin-based amplifier within the bandwidth
\begin{equation}
\Upsilon_{\rm{amp}}(\nu,\nu_{0})=\Upsilon_{\rm{amp}}(\nu_0,\nu_{0}) \times \dfrac{\sqrt{(\nu-\nu_{0})^2+(\Lambda/2)^2}}{\Lambda/2},
\label{P1}
\end{equation}
The magnetic sensitivity of $^{87}$Rb magnetometer assisted with the $^{129}$Xe spin-based amplifier can be achieved over different bias fields.


\end{itemize}

\subsection{Axion-like dark matter search experiment}
\label{sec5B}

\subsubsection{The timing of axion-like dark matter searches}

\begin{figure*}[t]  
	\makeatletter
\centering
	\def\@captype{figure}
	\makeatother
	\includegraphics[scale=1.6]{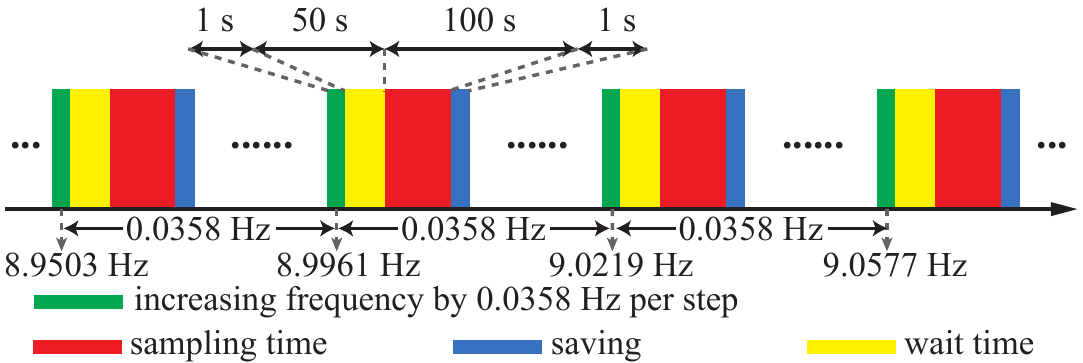}
	\caption{\textbf{The timing of axion-like dark matter search.} A single search consists of 50~s waiting time and 100~s sampling time. Before each search, the $^{129}$Xe Larmor frequency is increased by 0.0358~Hz per step by changing the bias field.}
	\label{SearchTiming}
\end{figure*}

The spin-based amplifier can operate with high amplification factor within the bandwidth of 0.052~Hz centered at $^{129}$Xe Larmor frequency,
as discussed in Sec.~\ref{sec2B}.
Due to the narrow bandwidth, the spin-based amplifier is only highly sensitive to a narrow range of axion-like particle masses. For example, the current 0.052~Hz bandwidth corresponds to 0.2~feV bandwidth of axion-like particle mass (see Eq.~\ref{J1}). To search for a broad range of axion mass, we can move the high-sensitivity bandwidth window by changing bias field, which can accordingly change the $^{129}$Xe Larmor frequency of the spin-based amplifier. The detailed procedure of our search experiment is shown in Fig.~\ref{SearchTiming}. In our search experiment, the scanning step of $^{129}$Xe Larmor frequency is 0.0358~Hz, which is slightly smaller than the spin-based amplifier bandwidth (i.e., 0.052~Hz).
In principle, we could choose a smaller scanning step, but it would take more time to complete the entire search.
The entire $^{129}$Xe Larmor frequency range is from 2~Hz to 180~Hz, corresponding to axion-like particle masses from 8.3~feV to 0.74~peV. In each search, we take 100~s of data and calculate the detection threshold of axion-like dark matter (see below). Last, we combine the all independent search results to achieve the entire detection threshold from 2~Hz to 180~Hz. The detailed data processing is described in the following. As an initial search, we also report experimental search for 100 probed axion-like particle mass windows in the range of 10~feV to 1~peV, where each result is from 5~h of data.

\begin{figure*}[t]  
	\makeatletter
\centering
	\def\@captype{figure}
	\makeatother
	\includegraphics[scale=1.45]{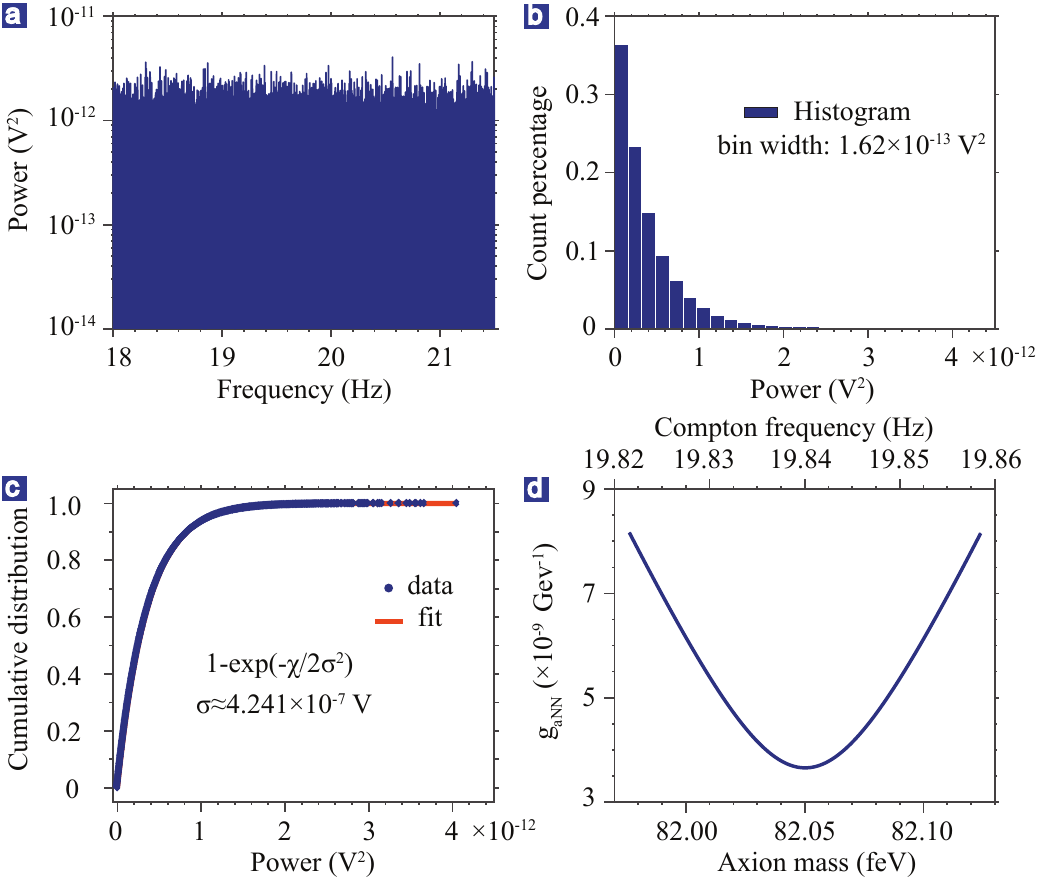}
	\caption{\textbf{Data analysis of axion-like dark matter search}. (a) A small window of the experimental power spectrum. In this situation, the bias field is set to 1681.3~nT and the corresponding $^{129}$Xe Larmor frequency is $\nu_0 \approx 19.839$~Hz. The searched axion mass window is centered at 82.05~feV with the width of 0.15~feV. (b) White-noise power histogram, which can be modelled as a chi-squared distribution with 2 degrees of freedom. (c) Cumulative distribution of chi-square distribution and fit to the exponential distribution $1-e^{-\chi/2\sigma^2}$. The fit yields a white-noise variance $\sigma \approx 4.241 \times 10^{-7}$~V. (d) Limits ($95\%$ confidence level) on the coupling strength of nucleons with the axion-like dark matter within a mass range centered at $m_a=82.05$~feV and with the width of 0.15~feV. The limits are ahieved by five hour measurement at $\nu_0 \approx 19.839$~Hz of the spin-based amplifier.}
	\label{AxionSearch}
\end{figure*}

\subsubsection{Detection threshold determination}

To define the detection threshold, we use the standard $p$-value hypothesis test. Detection threshold for a hypothesis test at a frequency $\nu$ is defined by the probability to find data which are less compatible than the observed $z_{0}$.
In the following, we use one of 5-h search data ($\nu_0 \approx 19.839$~Hz, corresponding to $m_a=82.05$~feV) as an example to explain our data procedure.
We calculate the power spectrum of the 5-h data. Figure~\ref{AxionSearch}$\textbf{a}$ shows the amplitude spectrum (the square root of the power spectral density). We analyze the power spectral density in $\pm 2$~Hz region centered at the corresponding Larmor frequency.
The histogram of the power spectral density in this frequency is shown in Fig.~\ref{AxionSearch}$\textbf{b}$ with the bin width of $1.62\times 10^{-13}$~V$^2$. The histogram can be modeled as the chi-squared distribution with two degrees of freedom. The corresponding cumulative distribution function (CDF) follows an exponential distribution~\cite{walck2007hand}, as shown in Fig.~\ref{AxionSearch}$\textbf{c}$. Specifically, the probability that the measured power $P(\nu)$ is smaller than $z_{0}$ is
\begin{equation}
\textrm{Pr}\{\textrm{P}(\nu)<z_{0}\}=1-e^{-z_{0}/2\sigma^{2}},
\label{O1}
\end{equation}
where $\sigma$ is the standard deviation. Hence, we can fit the experimental data to obtain the standard deviation $\sigma$. The red curve is the fit and agrees well with the experiment, yielding $\sigma \approx 4.2 \times 10^{-7}$~V. For 100~s sampling, we adopt the same data processing method.

We now define a signal power threshold $z_0$ such that, if at the probed frequency $\nu$, the power of a spectrum containing both the signal and noise, is greater than $z_0$, then peak has a $p_0$ probability of being induced by noise fluctuation,
\begin{equation}
p_0:=\textrm{Pr} \{\textrm{P}(\nu) > z_0\}=e^{-z_0/2\sigma^2}.
\end{equation}
Based on the above equation, we can achieve
\begin{equation}
z_0=-2 \sigma^2 \textrm{ln}(p_0).
\end{equation}
Thus, when $p_0=0.05$ (corresponding to 95$\%$ confidence level), we find the signal detection threshold is
\begin{equation}
z_0=5.99 \sigma^2.
\end{equation}
Because our analysis is done on the amplitude spectrum, we define the 95$\%$ confidence detection threshold from the amplitude spectrum as
\begin{equation}
A_{\textrm{th}}=2.44 \sigma.
\end{equation}
Here, we should note that the unit of $A_{\textrm{th}}$ is voltage and can be converted into magnetic field according to the calibration (2) in Sec.~\ref{sec5A}. Further, to obtain the constraint on the coupling parameter $\textrm{g}_{\rm{aNN}}$ (the coupling of axion-like particles to nucleons), we can use Eq.~\ref{J8} and $A_{\rm{th}}$ to achieve the corresponding $\textrm{g}_{\rm{aNN}}$.
For example, our experiment constrains the parameter space describing the coupling to axion-like dark matter, $|\textrm{g}_{\rm{aNN}}|< 3.5 \times 10^{-9}$~GeV$^{-1}$ at 82.05~feV axion mass.

In each search experiment (corresponding to a bias field), not only $\textrm{g}_{\rm{aNN}}$ can be constrained at the corresponding Larmor frequency, but also $\textrm{g}_{\rm{aNN}}$ can be constrained at near-resonance frequencies,
because the spin-based amplifier still has a high sensitivity to axion-like signals whose oscillating frequencies are within the amplifier bandwidth.
We have

\begin{equation}
\textrm{g}_{\textrm{aNN}} (\nu,\nu_{0})=\textrm{g}_{\textrm{aNN}}(\nu_0,\nu_{0}) \times \dfrac{\sqrt{(\nu-\nu_{0})^2+(\Lambda/2)^2}}{\Lambda/2},
\label{gann_band}
\end{equation}
where $\textrm{g}_{\textrm{aNN}}(\nu_0,\nu_{0})$ is the axion-nucleon coupling limit at the Larmor frequency $\nu_0$ and $\textrm{g}_{\textrm{aNN}} (\nu,\nu_{0})$ is the axion-nucleon limit at the near-resonance frequency.
Figure~\ref{AxionSearch}$\textbf{d}$ provides the constraint limit on $\textrm{g}_{\rm{aNN}}$ over the spin-based amplifier bandwidth. After obtaining the threshold for the spin-based amplifier bandwidth range, we scan the bias field to obtain the threshold of axion-like dark matter from 2~Hz and 180~Hz. The limit for $\textrm{g}_{\textrm{aNN}}$ is shown in Fig.~4\textbf{b} of the main text.

\section{Constraint plots for other possible bosonic dark matter fields}

\begin{figure*}[t]  
	\makeatletter
    \centering
	\def\@captype{figure}
	\makeatother
	\includegraphics[scale=2]{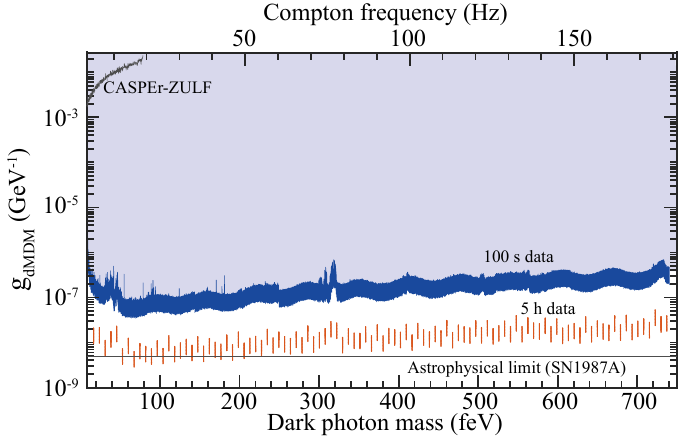}
	\caption{\textbf{Results of axion-like dark matter search}. Limits on dark photon–nucleon dMDM coupling in the mass range from 8.3~feV to 744~feV. The blue-shade is excluded by our measurements (100~s for each run) at the $95\%$ confidence level. The red lines show our advanced sensitivity (five hours for each run) for dark photon-nucleon interactions at 100 probed dark-photon masses with a window width of 0.15~feV. The grey line shows the limit given by the CASPEr-ZULF experiment~\cite{garcon2019constraints}. The horizontal black line shows the astrophysical limit from supernova SN1987A cooling~\cite{vysotsskii1978some, raffelt2008astrophysical}.}
	\label{gdmdm}
\end{figure*}

This section presents three other possible couplings between bosonic dark matter fields and nuclear spins~\cite{pospelov2013detecting, graham2015experimental, graham2018spin, wu2019search, garcon2019constraints}: the quadratic wind coupling with the axions, the couplings to dark electric field and magnetic field mediated by spin-1 bosons such as dark photons~\cite{wu2019search,garcon2019constraints}. The Hamiltonian of the three couplings can be written as
\begin{align}
	\label{T1}
	&\mathcal{H}_{\textrm{quad}} \approx 2\textrm{g}_{\textrm{quad}}^2\hbar^2 c^2 \dfrac{\rho_{\textrm{DM}}}{2\pi\nu_{a}}\sin(2\pi \nu_a t) \bm{v} \cdot \textbf{I}_{\rm{N}},\\
	\label{T2}
	&\mathcal{H}_{\rm{dMDM}}
    \approx \textrm{g}_{\rm{dMDM}} \sqrt{2 \hbar^3 c \rho_{\rm{DM}}} \sin(2\pi \nu_{a}t) \bm{v} \cdot \textbf{I}_{\rm{N}},\\
	\label{T3}
	&\mathcal{H}_{\rm{dEDM}}
	\approx \textrm{g}_{\rm{dEDM}} \sqrt{2 \hbar^3 c \rho_{\rm{DM}}} \sin(2\pi \nu_{a}t) \bm{v} \cdot \textbf{I}_{\rm{N}} /|\bm{v}|,
\end{align}
where $\textrm{g}_{\textrm{quad}}$ is the quadratic coupling strength between axions and nuclear spins, $\textrm{g}_{\rm{dMDM}} $ and $\textrm{g}_{\rm{dEDM}}$ respectively parameterizes the coupling strengths between nuclear spins with the dark magnetic fields and electric fields. $\bm{v}$ is taken as the same $\bm{v}$  in Sec.~\ref{sec4}. Based on the same analysis and calculation in Sec.~\ref{sec4}, we can derive the effective magnetic field induced by the bosonic dark matter fields

\begin{align}
	\nonumber
	\textbf{B}_{\textrm{quad}}[\textrm{T}] &\approx 10^{-3} \dfrac{\textrm{g}_{\rm{quad}}^2}{ \gamma_{n}} 2 \hbar c^3 \dfrac{\rho_{\rm{DM}}}{2\pi\nu_a}\sin(2\pi \nu_{a}t) \times \dfrac {\bm{v}}{|\bm{v}|},\\
	\label{T4}
	&\approx 2.9 \cdot 10^{-4} \dfrac{\rm{g}_{\rm{quad}}^2[(\rm{GeV}^{-1})^2]}{\rm{g}_{n}}\dfrac{\sin(2\pi \nu_{a}t)}{2\pi\nu_a} \times \sqrt{1-3/4 \cos^{2}(\Omega t- \eta)},\\
	\nonumber
	\textbf{B}_{\textrm{dMDM}} [\textrm{T}]&\approx 10^{-3} \dfrac{\textrm{g}_{\rm{dMDM}}}{\hbar \gamma_{n}} \sqrt{2 \hbar^{3} c^3 \rho_{\rm{DM}}}\sin(2\pi \nu_{a}t) \times \dfrac{\bm{v}}{|\bm{v}|},\\
	\label{T5}
	&\approx 10^{-7} \dfrac{\rm{g}_{\rm{dMDM}}[\rm{GeV}^{-1}]}{\rm{g}_{n}}\sin(2\pi \nu_{a}t) \times \sqrt{1-3/4 \cos^{2}(\Omega t- \eta)},\\
	\nonumber
	\textbf{B}_{\textrm{dEDM}} [\textrm{T}]&\approx  \dfrac{\textrm{g}_{\rm{dEDM}}}{\hbar \gamma_{n}} \sqrt{2 \hbar^{3} c \rho_{\rm{DM}}}\sin(2\pi \nu_{a}t) \times \dfrac{\bm{v}}{|\bm{v}|},\\
	\label{T6}
    &\approx 10^{-4} \dfrac{\rm{g}_{\rm{dEDM}}[\rm{GeV}^{-1}]}{\rm{g}_{n}}\sin(2\pi \nu_{a}t) \times \sqrt{1-3/4 \cos^{2}(\Omega t- \eta)}.
\end{align}
Thus, based on the same procedures to derive the constraints level of the axion-nucleon coupling (see Sec.~\ref{sec5}), the constraint plots for the couplings between the nuclear spins and the bosonic dark matter fields are shown in Fig.~\ref{gdmdm} ($\textrm{g}_{\textrm{dMDM}}$), Fig.~4\textbf{c} ($\textrm{g}_{\textrm{dEDM}}$) and Fig.~4\textbf{d} ($\textrm{g}_{\textrm{quad}}$) of the main text.

\section{Projected sensitivity reach}
\label{sec2C}

Our experimental results demonstrate the feasibility of using spin-based amplifier to search for axion-like dark matter.
We anticipate that the experimental sensitivity to axion-like dark matter can be further improved by several orders of magnitude.
The main factors limiting the sensitivity of our search for axion-like dark matter are:
(1) the amplification factor of the spin-based amplifier,
and (2) the magnetic sensitivity of the alkali-metal magnetometer itself.
Considering this,
we plan to employ $^3$He-K systems as a spin-based amplifier,
because $^3$He spins have much longer coherence time ($T_2 \sim 1000$~s) and larger gyromagnetic ratio than those of $^{129}$Xe ($T_2 \sim 20$~s in this work).
In the following, we estimate the corresponding amplification factor and axion-like dark matter search sensitivity.

We theoretically consider the amplification effect on $^3$He-K system.
Compared to $^{87}$Rb magnetometer, K magnetometer has demonstrated higher sensitivity~\cite{kominis2003subfemtotesla, kornack2005nuclear} and a $^3$He-$^{39}$K spin-based amplifier can achieve larger amplification factor as stated below. Based on the Eq.~\ref{G3}, we predict the amplification factor and higher magnetic sensitivity of such a spin-based amplifier. To calculate the amplification factor, we need to calculate the term $\lambda M^{n} P^{n}_{0}$. Using the SI units, $\lambda M^{n} P^{n}_{0}$ can be expressed as~\cite{walker1997spin, kornack2005nuclear}
\begin{equation}
\begin{aligned}
\dfrac{2\mu_{0}}{3} \kappa_{0} M^{n} P_{0}^{n}=\dfrac{2\mu_{0}}{3} \kappa_{0}g \mu P_{0}^{n} [\rm{N}],
\label{B1}
\end{aligned}
\end{equation}
where $\kappa_{0}$ is the enhancement factor for the given alkali metal and noble gas pairs (for example, for $^3$He and K, $\kappa_{0} \approx 5.9$),
$g$ is the Land$\acute{e}$ factor for noble gas nucleus,
$P^{n}_{0}$ is the polarization of noble gas nucleus,
$[\rm{N}]$ is the number density of the noble gas,
and $\mu=1.07 \times 10^{-26}~\textrm{J/T}$ is the magnetic moments of $^3$He nucleus.
We use Eq.~\ref{B1} to estimate the $^3$He effective magnetic field and numerically calculate the amplification factor.
Based on the demonstrated experimental parameters in ref.~\cite{kornack2005nuclear}, $P_{0}^{n} \approx 0.02$ is the typical polarization of the $^3$He gas,
$[\rm{He}] \approx 1.88 \times 10^{26}~\textrm{m}^{-3}$ corresponds to the number density with 7~amg $^3$He.
We obtain $\lambda M^n P_0^n \approx 2.0~\textrm{mG}$, which is in good agreement with 2.2~mG demonstrated in ref.~\cite{kornack2005nuclear}.
According to Eq.~\ref{eta},
the amplification factor of $^3$He-K spin-based amplifier is estimated to be $\eta \approx 10^{4}$.
Compared to $^{129}$Xe-Rb systems,
the amplification factor of $^3$He-K systems yields about two orders of magnitude improvement.

Based on the above discussion, one can evaluate the experimental sensitivity to axion-like dark matter.
We still need to know the sensitivity of the K magnetometer itself.
The $^3$He-K system has five orders smaller spin-destruction cross section than that of $^{129}$Xe-Rb sytems,
and thus the K magnetometer can still achieve a femtotesla sensitivity $\sim$$ 1~\textrm{fT} / \textrm{Hz}^{1/2}$, as demonstrated in ref.~\cite{kornack2005nuclear};
thus, the magnetic sensitivity based on $^3$He spin amplifier could probably reach $1~\textrm{aT} / \textrm{Hz}^{1/2}$ within the amplifier bandwidth.
We note that the experimental sensitivity to $\rm{g}_{\rm{aNN}}$ scales as $t^{-1/2}$ as a function of measurement time $t$.
When the measurement time is five hours,
the search sensitivity of $|\textrm{g}_{\textrm{aNN}}|$ can reach $ \sim$$ 10^{-13}~\textrm{GeV}^{-1}$,
which improves over the astrophysical limit (from supernova SN1987A cooling~\cite{vysotsskii1978some, raffelt2008astrophysical}) by at least four orders of magnitude.
For dark photon-nucleon interactions,
the experimental sensitivity could reach $\rm{g}_{\rm{dEDM}} \sim 10^{-17}$~GeV$^{-1}$ and $\rm{g}_{\rm{dMDM}} \sim 10^{-13}$~GeV$^{-1}$,
which are far beyond the astrophysical limits.
Using such a spin-based amplifier,
it is possible to explore new physics beyond the standard model.

If an alarm signal is observed, experiment would involve re-scans of the candidate frequencies to verify if they are statistical deviations or real axion-like signals.
The re-scans would be particularly time-demanding if the experiment is based on long-term search data.
Alternatively,
we plan to build a network of synchronized spin-based amplifiers to differentiate false alarms from true detection events.
Because the apparatus used in our experiment is small-scale and inexpensive,
we could quickly build such a network comprising of, for example, ten spin-based amplifiers.
Each sensor in the network independently performs the search for axion-like signals in the same frequency range from 1 Hz to 200~Hz.
Using the network,
claims of detecting axion-like dark matter could be made for candidates that are all above the thresholds for every sensor measurement at the searched frequency.
This leads to a significant decrease in false-alarm probability.
For example, with a network of ten spin-based amplifiers, the false-alarm probability becomes $p_0^{10} \approx 10^{-13}$ and the false-alarm rate is $\sim 10^{-7}$ alarms per experiment.
In addition to reducing false-alarm rate,
such a sensor network is promising to compose an exotic field telescope array for multi-messenger astronomy, as recently proposed~\cite{dailey2020quantum, pospelov2013detecting},
and address the stochastic fluctuations of bosonic dark matter~\cite{centers2019stochastic}.



\section{Testing the data analysis procedure with simulated axion signals}

\begin{figure*}[t]  
	\makeatletter
    \centering
	\def\@captype{figure}
	\makeatother
	\includegraphics[scale=1.14]{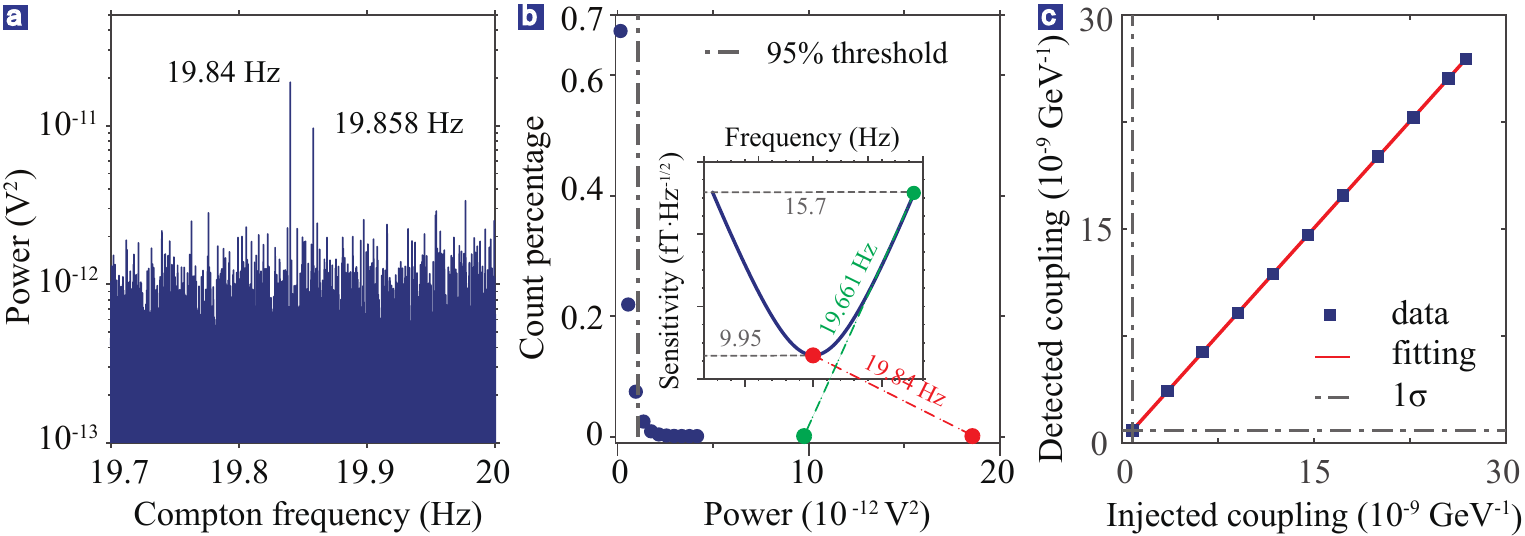}
	\caption{\textbf{Injecting simulated axion-like dark matter signals into the experimental data}. (a) A small window of the experimental power spectrum, with two numerically-generated synthetic axion-like signals added. The Compton frequencies of such axion-like signals are set to 19.84~Hz and 19.858~Hz and the axion-nucleon coupling strength is set to $\textrm{g}_{\rm{aNN}}$=25.18$\times 10^{-9}$~GeV$^{-1}$. The resonant frequency of the spin-based amplifier is 19.84~Hz. There are two visible axion-like signals at their corresponding Compton frequencies. The peak amplitude at resonant 19.84~Hz is larger than that at near-resonant 19.858~Hz. (b) The histogram of the power spectral data within the bin shown in (a). The vertical dashed line marks the 5.99$\sigma^{2}$ threshold ($95\%$ confidence level) for an event candidate. The injected synthetic axion-like signal was detected with 20$\sigma$ significance. The inset shows that the magnetic sensitivity is about 9.95~fT/Hz$^{1/2}$ at the resonant frequency and 15.7~fT/Hz$^{1/2}$ at the near-resonant frequency. (c) Verification of detection and coupling strength recovery of synthetic axion-like signals, with coupling strengths between $1.26 \times 10^{-9}$GeV$^{-1}$ and $25.18 \times 10^{-8}$~GeV$^{-1}$. The coupling strengths recovered from detected signals are shown as blue squares, and the coupling matching the injected signal is shown as the red line.}
	\label{Simulation}
\end{figure*}

Our data analysis procedure is tested by adding simulated axion-like dark matter signals into our experimental data.
As a demonstration, an on-resonance axion-like signal and a signal with a small frequency deviation from resonance frequency (corresponding to near-resonant case but still within the bandwidth) are injected into experimental data.
For example, we add two simulated axion-like signals at the Compton frequency of 19.84~Hz (on resonance) and 19.858~Hz (near-resonance) and
with the coupling strength $\textrm{g}_{\rm{aNN}}$=25.18$\times 10^{-9}$~GeV$^{-1}$.
These simulated signals can generate an effective magnetic field of 1.63$\times 10^{-3}$~pT.
The corresponding power spectrum is shown in Fig.~\ref{Simulation}$\textbf{a}$,
where the axion-like signals are clearly visible at 19.84~Hz and 19.858~Hz.
As shown in Fig.~\ref{Simulation}$\textbf{b}$,
there are two notable points (blue and red circles) that are significantly above the 95$\%$ threshold in the the histogram.
Based on the response of the spin-based amplifier,
we recover the effective magnetic field as 1.63$\times 10^{-3}$~pT and the coupling strength $\rm{g}_{\rm{aNN}}$=25.18$\times 10^{-9}$~GeV$^{-1}$, which agrees well with the injected simulated axion-like signals.
It indicates that our experiment can distinguish the simulated axion-like dark matter signal within the spin-based amplifier bandwidth.
Furthermore, we inject simulated signals from 1$\sigma$ to 20$\sigma$ significance corresponding coupling strength from 1.26$\times 10^{-9}$~GeV$^{-1}$ to 25.18$\times 10^{-9}$~GeV$^{-1}$.
The recovery axion-nucleon coupling strength are shown with blue squares in Fig.~\ref{Simulation}$\textbf{c}$, which are fitted with a linear fit (red line). The slope of the linear fit is 0.996, yielding that the couplings match the injected signals.

\bibliographystyle{naturemag}
\bibliography{supplementrefs}